\documentclass[aps,prd,reprint,nofootinbib,eqsecnum,showpacs]{revtex4-1}
\usepackage{amssymb,amsmath,amsfonts}                     
\usepackage{mathtools}                                    
\usepackage{bm}                                           
\usepackage{graphicx}                                     
\usepackage{subfigure}                                    
\usepackage{hyperref}                                     
\usepackage[capitalise]{cleveref}                         
\usepackage{dcolumn}                                      
\usepackage{xstring}                                      
\hypersetup{
	colorlinks=true,
	linkcolor=blue,
	citecolor=blue,
	urlcolor=blue
}
\newcommand{\crefrangeconjunction}{--}                    
\newcommand{\creflastconjunction}{, and\nobreakspace}     
\crefname{section}{Sec.}{Secs.}                           
\Crefname{section}{Section}{Sections}                     
\crefformat{subequations}{Eqs.~(#2#1#3)}                  
\Crefformat{subequations}{Equations~(#2#1#3)}             
\crefname{appsec}{Appendix}{Appendices}                   
\Crefname{appsec}{Appendix}{Appendices}                   
\newcommand{\refcite}[1]{
	\begingroup
		\def\tempx{0}%
		\StrCount{#1}{,}[\tempx]%
		\ifnum\tempx > 0
			Refs.~\cite{#1}%
		\else
			Ref.~\cite{#1}%
		\fi
	\endgroup
}
\newcommand{\Refcite}[1]{
	\begingroup
		\def\tempx{0}%
		\StrCount{#1}{,}[\tempx]%
		\ifnum\tempx > 0 
			References~\cite{#1}%
		\else
			Reference~\cite{#1}%
		\fi
	\endgroup
}
\newcommand{\der}[2]{\frac{\mathrm{d}#1}{\mathrm{d}#2}}   

\newcommand{\parder}[2]{\frac{\partial#1}{\partial#2}}    
\newcommand{\nparder}[3]{\frac{\partial^#1#2}{
	\partial#3^#1}}
\begin{document}


\title{Saturation of the \texorpdfstring{$f$}{f}-mode instability in neutron stars \texorpdfstring{\\}{:} I. Theoretical framework}

\author{Pantelis Pnigouras}
	\affiliation{Theoretical Astrophysics, IAAT, Eberhard-Karls University of T\"ubingen, 72076 T\"ubingen, Germany}
\author{Kostas D. Kokkotas}
	\affiliation{Theoretical Astrophysics, IAAT, Eberhard-Karls University of T\"ubingen, 72076 T\"ubingen, Germany}

\date{\today}


\begin{abstract}
	The basic formulation describing quadratic mode coupling in rotating Newtonian stars is presented, focusing on polar modes. Due to the Chandrasekhar-Friedman-Schutz mechanism, the $f$-mode (fundamental oscillation) is driven unstable by the emission of gravitational waves. If the star falls inside the so-called instability window, the mode's amplitude grows exponentially, until it is halted by nonlinear effects. Quadratic perturbations form three-mode networks inside the star, which evolve as coupled oscillators, exchanging energy. Coupling of the unstable $f$-mode to other (stable) modes can lead to a parametric resonance and the subsequent saturation of its amplitude, thus suppressing the instability. The saturation point determines the amplitude of the gravitational-wave signal obtained from an individual source, as well as the evolutionary path of the latter inside the instability window.
\end{abstract}



\maketitle


\section{Introduction} \label{sec:Introduction}

With a mass of the order of the solar mass and a radius of about 10 km, neutron stars constitute nature's high-energy laboratories, from which the behavior of matter at such extreme conditions could be deduced. The neutron star equation of state is yet to be determined and remains one of the most significant questions in astrophysics. The serendipitous discovery of the first pulsar by Hewish and Bell in 1967 signified the onset of neutron star astronomy, which has provided some constraints for the masses, the radii, and the rotation periods of neutron stars.

Nevertheless, these observations are not enough to infer the equation of state.  A method that could be used to further probe neutron stars is asteroseismology, namely, the study of stellar oscillations \cite{UnnoEtAl1989,AertsEtAl2010}. Especially after the realization that stellar oscillations can be driven unstable by the emission of gravitational radiation \cite{Chandrasekhar1970,FriedmanSchutz1978,*FriedmanSchutz1978b}, the field of \textit{gravitational wave asteroseismology} was developed rapidly; detection of gravitational waves from nonradial stellar oscillations could provide information about the neutron star interior \cite{AnderssonKokkotas1996,AnderssonKokkotas1998,KokkotasEtAl2001,BenharEtAl2004,GaertigKokkotas2011}.

The Chandrasekhar-Friedman-Schutz (CFS) instability, however, grows on long time scales and, to make things worse, it is suppressed by viscosity \cite{IpserLindblom1990,IpserLindblom1991}. For the $f$-modes, which are the fundamental oscillations of the star and the best gravitational wave emitters, this leaves only a small portion of the parameter space, where the instability is active. In the late 1990s, it was realized that another class of oscillations, the $r$-modes, is unstable for a much larger parameter range \cite{Andersson1998,FriedmanMorsink1998,LindblomEtAl1998,OwenEtAl1998,AnderssonKokkotas2001}. The $r$-modes are related to horizontal motions of the fluid, much like Rossby waves in the Earth's atmosphere and oceans, and exist only in rotating stars \cite{PapaloizouPringle1978}. Moreover, they have shorter growth times, compared to the $f$-modes. As a result, the $r$-mode instability was considered as the most promising gravitational wave source.

Consequent studies on the $r$-mode instability naturally raised the question of the maximum amplitude that the oscillation can attain, before it is halted by nonlinear effects. Coupling of the unstable $r$-mode to other modes of the star can work as an energy drain and saturate the instability. The results of these studies were quite disappointing, from a gravitational-wave-detection point of view: the $r$-mode saturation amplitude is, in fact, quite lower than expected, or, at least, hoped \cite{SchenkEtAl2001,Morsink2002,ArrasEtAl2003,BrinkEtAl2004,*BrinkEtAl2004b,*BrinkEtAl2005}.

Determining the saturation amplitude of the unstable $r$-mode is also important for neutron star evolution. Whether the star is newborn or a member of a low-mass x-ray binary system (LMXB), its evolution depends on the value of the saturation amplitude \cite{Levin1999,BondarescuEtAl2007,*BondarescuEtAl2009}. When the star enters the instability region, it loses angular momentum, due to gravitational wave emission, which could possibly explain the upper limit in the observed neutron star rotational frequencies \cite{Friedman1983,LindblomEtAl1998,AnderssonEtAl1999,AnderssonEtAl1999b,BondarescuWasserman2013} (about 700 Hz \cite{HesselsEtAl2006}).

Even though the $r$-mode instability is active in a much larger part of the parameter space, the $f$-mode instability could still be significant, especially for newborn neutron stars. Furthermore, the fact that the $r$-mode saturation amplitude is not expected to be high renders the study of the $f$-mode quite important: if the $f$-mode is not saturated at such low amplitudes, then it could be a possible gravitational wave source and, thus, provide much information about the neutron star equation of state. Up until now, the evolution of the $f$-mode instability in the nonlinear regime has been performed only via hydrodynamic simulations \cite{ShibataKarino2004,OuEtAl2004,KastaunEtAl2010}. However, since the growth time of the instability is, in general, quite long, it is very hard for nonlinear simulations to track the mode evolution for such a long time.

Recent work \cite{PassamontiEtAl2013} suggests that, should the $f$-mode saturate at reasonably high amplitudes, the gravitational wave signal from a source in the Virgo cluster, undergoing the $f$-mode instability, could be detectable by the Einstein Telescope. A more promising source is related to supramassive configurations (exceeding the maximum mass of a nonrotating star), which could be the outcome of a neutron star merger. Such stars would be stable only for rotation rates close to the Kepler limit (mass-shedding limit). The $f$-mode instability is expected to grow really quickly in these objects and the gravitational wave signal could even reach the sensitivity of Advanced LIGO, with a quite promising event rate \cite{DonevaEtAl2015}.

As opposed to the $r$-mode, where the oscillation comprises horizontal fluid motions, the $f$-mode is dominated by a radial component and large-scale density variations, which makes it a more efficient gravitational wave emitter. However, the so-called instability window is much smaller for the $f$-mode. This is a region in the ``temperature-rotation rate'' plane, where the instability is not suppressed by viscous effects. By expanding a perturbation in its multipole moments (described by the spherical harmonics $Y_l^m$) we see that higher multipoles become unstable at lower rotation rates. On the other hand, lower multipoles emit gravitational waves more efficiently, but might not become unstable at all. The instability window of the $l=m=2$ $f$-mode is significant only for models with quite stiff equations of state, whereas $l=m=3$ and $4$ $f$-modes have larger windows, but might not grow very fast.

Applying the same methodology as for the $r$-mode, we can determine the amplitude at which the $f$-mode instability is saturated by nonlinear effects. This work has been divided into two parts. In the first part, included in the present paper, we will present the theoretical framework of the problem. Its application to various stellar models will be presented in a subsequent paper.

The paper is organized as follows: in \cref{sec:The oscillation modes---Linear perturbation scheme} we present the formalism that gives rise to the various oscillation modes in the star, using linear perturbations. We discuss the method with which one can acquire the oscillation spectrum in the nonrotating limit, and then we add rotation in a perturbative way (slow-rotation approximation) and present its main implications. In \cref{sec:The f-mode instability} we give a short overview of the CFS instability and how it is manifested in the $f$-mode. In \cref{sec:Mode coupling---Quadratic perturbation scheme} we review the formalism which describes quadratic perturbations and derive the conditions under which coupled-mode networks can arise. Furthermore, these networks are subjected to a stability analysis, which determines whether saturation can be achieved by the system or not. Derivations of  several formulas in this section are addressed in Appendices. In \cref{sec:Derivation of the equations of motion} 
we derive the equations of motion, including quadratic perturbations, whereas in \cref{sec:The coupling coefficient} we give the expression for the three-mode coupling coefficient. \Cref{sec:Study of a three-mode network with quadratic nonlinearities} contains a study of a coupled three-mode network, using the multiscale method, as well as the details of the stability analysis mentioned above. Finally, \cref{sec:Discussion} concludes the paper with some discussion.


\section{The oscillation modes---Linear perturbation scheme} \label{sec:The oscillation modes---Linear perturbation scheme}

Stellar oscillation modes can be divided in two general categories: \textit{polar} (or \textit{spheroidal}) modes and \textit{axial} (or \textit{toroidal}) modes. Expanding the displacement vector field of an arbitrary perturbation in vector spherical harmonics, we get
\begin{align}
	\bm{\xi}(r,\theta,\phi)&=\sum_l\sum_{m=-l}^l\left[W_l^m(r)Y_l^m(\theta,\phi)\bm{e}_r\right. \label{spherical harmonic decomposition} \\
	&+V_l^m(r)\nabla Y_l^m(\theta,\phi)\left.+U_l^m(r)\bm{e}_r\times\nabla Y_l^m(\theta,\phi)\right], \notag
\end{align}
where $(r,\theta,\phi)$ are the spherical polar coordinates, $(\bm{e}_r,\bm{e}_{\theta},\bm{e}_{\phi})$ is the orthonormal basis, and $Y_l^m$ are the spherical harmonics. Then
\begingroup
\renewcommand*{\arraystretch}{1.4}
\begin{equation*}
	\begin{array}{ll}
		\bullet\;\textrm{polar modes:} & U_l^m=0 \\
		\bullet\;\textrm{axial modes:} & V_l^m=W_l^m=0
	\end{array}
	\;\,\mathrm{as}\;\,\Omega\rightarrow 0,
\end{equation*}
\endgroup
$\Omega$ being the stellar rotation rate. $f$-modes, as well as $p$- (acoustic waves) and $g$-modes (gravity waves), are examples of polar modes. They constitute the ``regular'' mode spectrum of a star and have finite frequencies in the nonrotating limit. $r$-modes, on the other hand, are axial and become trivial in the nonrotating limit, where their frequencies vanish (for a detailed presentation of oscillation modes, cf. for instance, \refcite{UnnoEtAl1989,AertsEtAl2010}). The picture above slightly changes in the case of zero-buoyancy stars. $g$-modes, which are caused by the presence of buoyancy, become trivial too. The result of this ``mixture'' of trivial modes ($r$- and $g$-modes) is another class of modes, called \textit{hybrid} modes, which have both polar and axial components in the nonrotating limit. In the special case where $l=m$ one obtains the ``classical'' $r$-modes, which are purely axial \cite{LockitchFriedman1999}.

Assuming a star which is uniformly rotating with an angular velocity $\Omega$, the fluid equations, in the frame rotating with the star, are
\begin{gather}
	\parder{\rho}{t}+\nabla\cdot(\rho\bm{v})=0, \label{continuity} \\
	\parder{\bm{v}}{t}+(\bm{v}\cdot\nabla)\bm{v}+2\bm{\Omega}\times\bm{v}+\bm{\Omega}\times(\bm{\Omega}\times\bm{r})=-\frac{\nabla p}{\rho}-\nabla\Phi, \label{Euler}
\end{gather}
and
\begin{equation}
	\nabla^2\Phi=4\pi G\rho, \label{Poisson}
\end{equation}
where $\rho$ is the density, $p$ the pressure, $\bm{v}$ the velocity, $\Phi$ the gravitational potential and $G$ the gravitational constant. The system above has to be supplemented with an equation of state $p=p(\rho,\mu)$, where $\mu$ usually corresponds to entropy or composition and depends on the density. By considering ``small'' perturbations imposed on the equilibrium state, these equations are written as
\begin{gather}
	\parder{\delta\rho}{t}+\nabla\cdot(\rho\delta\bm{v})=0, \label{continuity perturbed} \\
	\parder{\delta\bm{v}}{t}+2\bm{\Omega}\times\delta\bm{v}=-\frac{\nabla\delta p}{\rho}+\frac{\nabla p}{\rho^2}\delta\rho-\nabla\delta\Phi, \label{Euler perturbed} \\
	\nabla^2\delta\Phi=4\pi G\delta\rho, \label{Poisson perturbed}
\end{gather}
and
\begin{equation}
	\frac{\Delta p}{p}=\Gamma_1\frac{\Delta\rho}{\rho}+\left(\parder{\ln p}{\ln\mu}\right)_{\rho}\frac{\Delta\mu}{\mu}, \label{EoS perturbed}
\end{equation}
where
\begin{equation}
	\Gamma_1=\left(\parder{\ln p}{\ln\rho}\right)_{\mu}. \label{Gamma_1}
\end{equation}
In the equations above $\delta$ denotes a Eulerian perturbation and $\Delta$ corresponds to a Lagrangian perturbation. The former monitors changes in a particular point in space, whereas the latter refers to changes in a given fluid element. The two are related by $\Delta f=\delta f+(\bm{\xi}\cdot\nabla)f$, where $\bm{\xi}$ is the Lagrangian displacement of the fluid element \cite{UnnoEtAl1989,Lynden-BellOstriker1967}.

By definition, $\Delta\bm{v}=\mathrm{d}\bm{\xi}/\mathrm{d}t=\dot{\bm{\xi}}+(\bm{v}\cdot\nabla)\bm{\xi}$, but, since $\bm{v}=\bm{0}$ in the background, $\Delta\bm{v}=\dot{\bm{\xi}}=\delta\bm{v}$. Then, the perturbed Euler equation \eqref{Euler perturbed} can be written as \cite{Lynden-BellOstriker1967}
\begin{equation}
	\ddot{\bm{\xi}}+\bm{\mathcal{B}}(\dot{\bm{\xi}})+\bm{\mathcal{C}}(\bm{\xi})=\bm{0}, \label{equation of motion}
\end{equation}
where
\begin{equation}
	\bm{\mathcal{B}}(\bm{\xi})=2\bm{\Omega}\times\bm{\xi}, \label{operator B}
\end{equation}
and
\begin{equation}
	\bm{\mathcal{C}}(\bm{\xi})=\frac{\nabla\delta p}{\rho}-\frac{\nabla p}{\rho^2}\delta\rho+\nabla\delta\Phi. \label{operator C}
\end{equation}
Operator $\bm{\mathcal{C}}$ can be written in terms of $\bm{\xi}$ by using \cref{continuity perturbed,Poisson perturbed,EoS perturbed} to replace the perturbations $\delta\rho$, $\delta\Phi$, and $\delta p$, respectively (cf. for example, Sec. II~B in \refcite{SchenkEtAl2001}, or Sec. 2.1 in \refcite{Lynden-BellOstriker1967}).

Seeking solutions of the form $\bm{\xi}(\bm{r},t)=\bm{\xi}(\bm{r})e^{i\omega t}$, where $\omega$ denotes the frequency of a mode in the corotating frame, \cref{equation of motion} is written as
\begin{equation}
	-\omega^2\bm{\xi}+i\omega\bm{\mathcal{B}}(\bm{\xi})+\bm{\mathcal{C}}(\bm{\xi})=\bm{0}. \label{eigenvalue equation}
\end{equation}
This is the eigenvalue equation which needs to be solved, supplemented with the appropriate boundary conditions, in order to obtain the mode spectrum of the star.


\subsection{The nonrotating limit} \label{subsec:The nonrotating limit}

\Cref{eigenvalue equation} is simplified significantly in the absence of rotation, since operator $\bm{\mathcal{B}}$ vanishes. Then, according to \cref{spherical harmonic decomposition}, the displacement vector of a polar mode is
\begin{equation}
	\bm{\xi}(r,\theta,\phi)=\left(\xi_r(r),\xi_h(r)\parder{}{\theta},\xi_h(r)\frac{1}{\sin\theta}\parder{}{\phi}\right)Y_l^m(\theta,\phi), \label{polar mode eigenfunction}
\end{equation}
where $\xi_r$ and $\xi_h$ are the radial and horizontal components of $\bm{\xi}$, respectively. It should be noted that, since operator $\bm{\mathcal{C}}$ is Hermitian \cite{Lynden-BellOstriker1967}, the solutions to \cref{eigenvalue equation} (with vanishing $\bm{\mathcal{B}}$) are orthogonal, i.e.
\begin{equation}
	\langle\bm{\xi}_\alpha,\bm{\xi}_\beta\rangle\equiv\int\rho\,\bm{\xi}_\alpha^*\cdot\bm{\xi}_\beta\mathrm{d}^3\bm{r}=I_\alpha\delta_{\alpha\beta}, \label{mode orthogonality condition in the non-rotating limit}
\end{equation}
where the indices in $\bm{\xi}$ denote different solutions, $\delta_{\alpha\beta}$ is the Kronecker delta, and the star denotes complex conjugation. Since all perturbative quantities are functions of $\bm{\xi}$, they can all be expressed as
\begin{equation*}
	\delta f(r,\theta,\phi,t)=\delta f(r)Y_l^m(\theta,\phi)e^{i\omega t}.
\end{equation*}
Hence, a separation of variables is possible and the problem is reduced to calculating the radial dependence of the perturbation \cite{UnnoEtAl1989}.

\begin{figure}[b]
	\includegraphics[width=.48\textwidth,keepaspectratio]{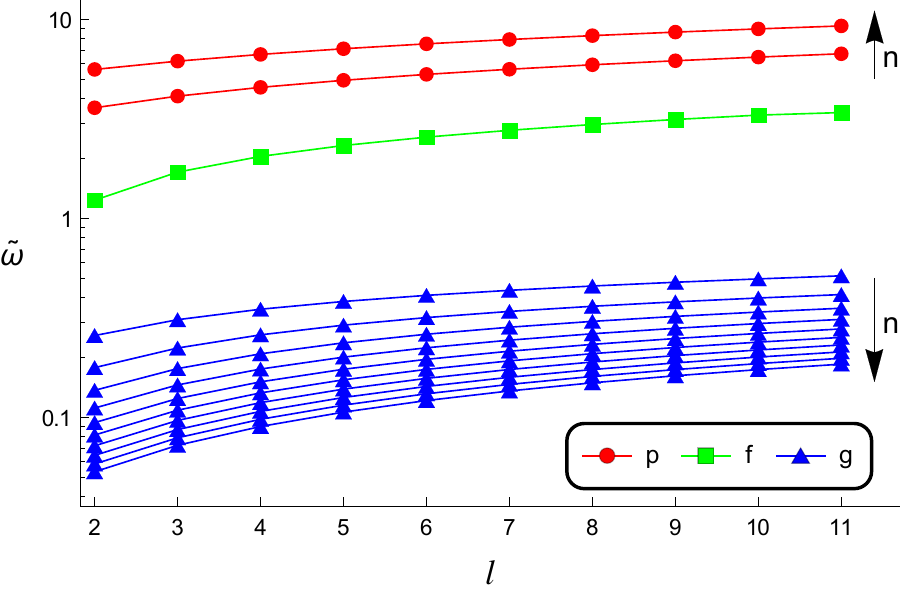}
	\caption{Polar mode spectrum of a star obeying a polytropic equation of state with $\Gamma=2$. The adiabatic exponent $\Gamma_1$ is equal to 2.1. Mode frequencies, which scale as $\tilde{\omega}=\omega/\sqrt{GM/R^3}$, are plotted against the mode degree $l$.}
	\label{fig:model 7 eigenfrequencies}
\end{figure}

A sample from the polar mode spectrum of a polytropic star is presented in \cref{fig:model 7 eigenfrequencies}. Each mode is generally described by three numbers: its overtone $n$, its degree $l$, and its order $m$. When rotation is absent, the mode frequencies do not depend on $m$ (see \cref{subsec:The slow-rotation approximation}). The $f$-mode ($n=0$) lies between its overtones ($n>0$), the high-frequency $p$-modes and the low-frequency $g$-modes. $g$-modes are pushed towards zero as the effects of buoyancy become less and less important, until they finally vanish for zero-buoyancy stars. Departure from the zero-buoyancy case can be a result of stratification (composition gradients) or deviations from isentropy (star with a finite temperature) \cite{Finn1986,*Finn1987}.

This behavior can be described by the so-called Schwarzschild discriminant, which is given by
\begin{equation*}
	A=\der{\ln\rho}{r}-\frac{1}{\Gamma_1}\der{\ln p}{r},
\end{equation*}
where $\Gamma_1$ is the adiabatic exponent, defined in \cref{Gamma_1}. If the star obeys a simple polytropic equation of state $p=K\rho^\Gamma$ (where $K$ and $\Gamma$ are constants), the Schwarzschild discriminant becomes
\begin{equation*}
	A=\frac{\Gamma_1-\Gamma}{\Gamma_1}\der{\ln\rho}{r}.
\end{equation*}
Then, if $\Gamma=\Gamma_1$ the star exhibits no convective phenomena (zero-buoyancy case). On the other hand, $\Gamma<\Gamma_1$ ($\Gamma>\Gamma_1$) denotes convective stability (instability), i.e. oscillatory (unstable) $g$-modes.

If the equation of state is described by the more general relation $p=p(\rho,\mu)$, the occurrence of convective phenomena is parametrized through $\mu$. The condition for the existence of $g$-modes is $\Delta\mu=0$ [cf. \cref{EoS perturbed}]. If $\mu$ corresponds to the composition, this condition means that the composition of a displaced fluid element is ``frozen''; weak interaction processes need more time than an oscillation period to restore $\beta$-equilibrium between the displaced fluid element and the surrounding matter. On the other hand, if $\mu$ corresponds to entropy, it means that the fluid displacement occurs adiabatically. The Schwarzschild discriminant, as a function of $\mu$, is given by
\begin{equation*}
	A=-\frac{1}{\Gamma_1}\left(\parder{\ln p}{\ln\mu}\right)_\rho\der{\ln\mu}{r}.
\end{equation*}


\subsection{The slow-rotation approximation} \label{subsec:The slow-rotation approximation}

Taking rotation into account, the situation changes significantly. The equilibrium configuration no longer exhibits spherical symmetry and an oscillation mode cannot be described by a single spherical harmonic. Typically, \cref{eigenvalue equation} has to be solved from scratch. However, rotation can also be introduced perturbatively, namely, by considering the effects of rotation to the various quantities as perturbations. Rotation affects polar modes in two ways. First, it lifts the $(2l+1)$-fold degeneracy in the eigenfrequency of each mode, by introducing a Zeeman-like splitting. The eigenfrequency now depends on both the degree $l$ and the order $m$, as opposed to the nonrotating limit, where it is degenerate in $m$. Second, rotation distorts the equilibrium structure of the star, which also changes the mode frequencies. An additional effect of rotation is, as discussed before, the appearance of a whole different class of modes, the inertial modes (like the $r$-mode), whose restoring force is the 
Coriolis force.

Mode splitting is already introduced as a first-order effect, whereas equilibrium distortion is a second-order effect. Higher-order effects also become important for large rotational velocities, but the analysis is quite cumbersome even at second order in $\Omega$. A third-order perturbation formalism was developed in \refcite{SoufiEtAl1998}, where an interesting case of near degeneracy was observed. Nevertheless, we stopped at quadratic perturbations in $\Omega$, keeping in mind that higher-order effects could be significant at the near-Kepler angular velocities that we are interested in.

\begin{figure}[b]
	\includegraphics[width=.48\textwidth,keepaspectratio]{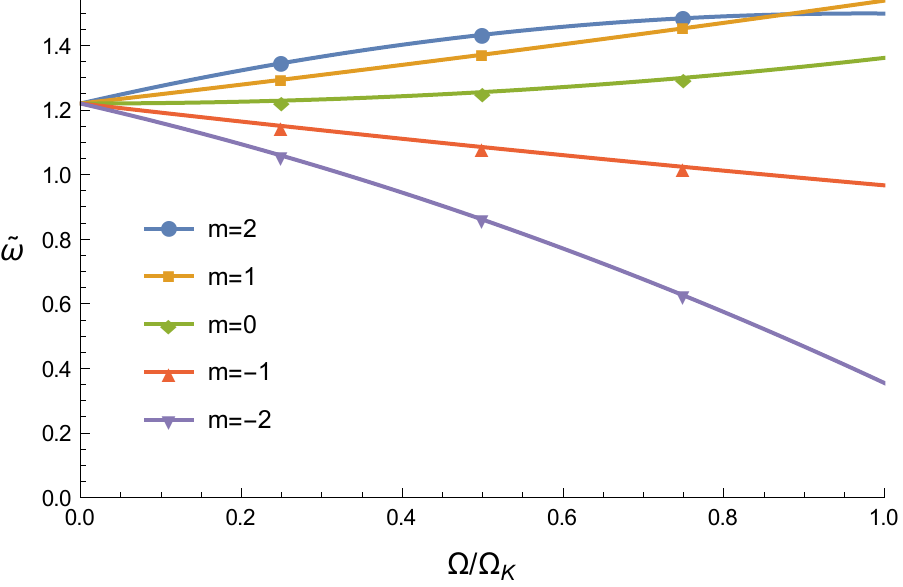}
	\caption{Eigenfrequency of the $l=2$ $f$-mode (in the corotating frame), as a function of the rotation rate $\Omega$. Each line corresponds to a different value of $m$. As in \cref{fig:model 7 eigenfrequencies}, a polytrope with $\Gamma=2$ and $\Gamma_1=2.1$ was used. The mode frequency scales as $\tilde{\omega}=\omega/\sqrt{GM/R^3}$, whereas the rotational velocity is normalized to the Kepler limit $\Omega_\mathrm{K}$.}
	\label{fig:model 7 l=2 f-mode}
\end{figure}

Eigenfrequencies, eigenfunctions, as well as equilibrium quantities, are expanded as
\begin{align*}
	\omega & =\omega_0+\omega_1(\Omega)+\omega_2(\Omega^2)+\mathcal{O}(\Omega^3), \\
	\bm{\xi} & =\bm{\xi}_0+\bm{\xi}_1(\Omega)+\bm{\xi}_2(\Omega^2)+\mathcal{O}(\Omega^3), \\
	\rho & =\rho_0+\rho_2(\Omega^2)+\mathcal{O}(\Omega^4).
\end{align*}
Substituting these in \cref{eigenvalue equation} and distinguishing between first- and second-order terms, we obtain \cite{SchenkEtAl2001}
\begin{equation}
	-\omega_0^2\bm{\xi}_1+\bm{\mathcal{C}}_0(\bm{\xi}_1)-2\omega_0\omega_1\bm{\xi}_0+i\omega_0\bm{\mathcal{B}}_1(\bm{\xi}_0)=\bm{0} \label{eigenvalue equation first-order}
\end{equation}
and
\begin{align}
	-\omega_0^2\bm{\xi}_2&+\bm{\mathcal{C}}_0(\bm{\xi}_2)-2\omega_0\omega_1\bm{\xi}_1+i\omega_0\bm{\mathcal{B}}_1(\bm{\xi}_1)-2\omega_0\omega_2\bm{\xi}_0 \notag \\
	&-\omega_1^2\bm{\xi}_0+i\omega_1\bm{\mathcal{B}}_1(\bm{\xi}_0)+\bm{\mathcal{C}}_2(\bm{\xi}_0)=\bm{0}, \label{eigenvalue equation second-order}
\end{align}
respectively. From the above, we find the $\mathcal{O}(\Omega)$ and $\mathcal{O}(\Omega^2)$ corrections to the eigenfrequencies. The first is rather simple and is given by
\begin{equation}
	\omega_1=mC_1\Omega, \label{eigenfrequency first-order correction}
\end{equation}
where
\begin{equation*}
	C_1=\frac{\int[2\xi_r\xi_h+\xi_h^2]\rho r^2\mathrm{d}r}{\int[\xi_r^2+l(l+1)\xi_h^2]\rho r^2\mathrm{d}r}.
\end{equation*}
The second is more complicated and has the general form \cite{Saio1981}
\begin{equation}
	\omega_2=C_2\Omega^2=(X+m^2Y)\Omega^2,
\end{equation}
where $X$ and $Y$ include corrections due to the distortion of the equilibrium and due to the effects of the Coriolis force. The effect of rotation on the mode eigenfrequencies (up to second order) can be seen in \cref{fig:model 7 l=2 f-mode}.

As for the eigenfunctions, rotation couples polar modes to axial modes, as well as other polar modes. This means that a mode cannot be any more described by a single spherical harmonic, which makes the situation more complicated. Since operator $\bm{\mathcal{B}}$ is nonvanishing in this case, the solutions to \cref{eigenvalue equation} do not obey the orthogonality relation \eqref{mode orthogonality condition in the non-rotating limit}. Instead, they satisfy a modified orthogonality condition, given by\footnote{Note that \refcite{SchenkEtAl2001} uses a different ansatz for $\bm{\xi}(\bm{r},t)$, i.e. $\bm{\xi}(\bm{r},t)=\bm{\xi}(\bm{r})e^{-i\omega t}$, hence the sign difference in the second term.} \cite{SchenkEtAl2001}
\begin{equation}
	(\omega_\alpha+\omega_\beta)\langle\bm{\xi}_\alpha,\bm{\xi}_\beta\rangle-\langle\bm{\xi}_\alpha,i\bm{\mathcal{B}}(\bm{\xi}_\beta)\rangle=b_\alpha\delta_{\alpha\beta}. \label{mode orthogonality condition}
\end{equation} 


\section{The \texorpdfstring{$f$}{f}-mode instability} \label{sec:The f-mode instability}

As it was discovered by Chandrasekhar \cite{Chandrasekhar1970} and rigorously proven by Friedman and Schutz \cite{FriedmanSchutz1978,*FriedmanSchutz1978b}, oscillation modes can be driven unstable by the emission of gravitational radiation, if the star is rotating rapidly enough. Every mode can be thought of as having a prograde (denoted by $-|m|$) and retrograde (denoted by $|m|$) component. Should the star rotate sufficiently fast, it can drag the retrograde component towards the direction of rotation, making it appear as prograde to a distant observer. Emission of gravitational waves by the perturbation can then act as a driving mechanism, increasing the mode energy. This can be seen by the standard multipole expansion of the power radiated in the form of gravitational waves (GW) \cite{Thorne1980}
\begin{equation}
	\left(\der{E}{t}\right)_\mathrm{GW}\hspace{-.3cm}=-\sum_{l_\mathrm{min}}^\infty N_l\,\omega\left(\omega-m\Omega\right)^{2l+1}\left(|\delta D_l^m|^2+|\delta J_l^m|^2\right). \label{GW energy rate}
\end{equation} 
As one can see, the power emitted is negative (gravitational radiation damps the mode), unless $\omega(\omega-m\Omega)<0$, in which case the energy of the mode is increased. The onset of the instability occurs when $\omega/m=\Omega$, namely when the \emph{pattern speed} of the mode matches the angular velocity of the star. The angular velocity at which this happens is usually called \emph{critical}. Alternatively, $\omega-m\Omega$ can be thought of as the mode frequency, measured in an inertial frame ($\omega$ is the corotating frame frequency). Then, the instability sets in at the point when the inertial-frame frequency changes sign.

In \cref{GW energy rate}, $N_l$ is a constant given by
\begin{equation}
	N_l=\frac{4\pi G}{c^{2l+1}}\frac{(l+1)(l+2)}{l(l-1)\left[(2l+1)!!\right]^2} \label{N_l}
\end{equation}
($c$ being the speed of light), whereas $\delta D_l^m$ and $\delta J_l^m$ denote the mass and current multipole moments, respectively. The $f$-mode radiates mainly via the former,\footnote{Current multipole moments become significant in the case of the $r$-modes (cf. for example, \refcite{LindblomEtAl1998}).} which are given by
\begin{equation}
	\delta D_l^m=\int r^l \delta\rho\, Y_l^{*m}\mathrm{d}^3\bm{r}. \label{mass multipole moments}
\end{equation}
Finally, the lower limit of the sum is given by $l_\mathrm{min}=\max (2,|m|)$.

Depending on the equation of state, all the $l=m$ $f$-modes can become unstable. However, various dissipation mechanisms are expected to act against the CFS instability. Responsible for the dissipation of the $f$-mode are mainly bulk and shear viscosity (BV and SV), and their contributions are given by \cite{IpserLindblom1991}
\begin{equation}
	\left(\der{E}{t}\right)_\mathrm{BV}=-\int\zeta\delta\sigma\delta\sigma^*\mathrm{d}^3\bm{r} \label{BV energy rate}
\end{equation}
and
\begin{equation}
	\left(\der{E}{t}\right)_\mathrm{SV}=-\int 2\eta\,\delta\sigma^{ij}\delta\sigma^*_{ij}\mathrm{d}^3\bm{r}, \label{SV energy rate}
\end{equation}
respectively. Here, $\delta\sigma^{ij}$ is the stress tensor and is given, in terms of the velocity perturbations, by
\begin{gather}
	\delta\sigma^{ij}=\frac{1}{2}\left(\nabla^i\delta v^j+\nabla^j\delta v^i-\frac{2}{3}g^{ij}\delta\sigma\right), \label{stress tensor} \\
	\delta\sigma=\nabla_i\delta v^i, \label{stress scalar}
\end{gather}
$g_{ij}$ being the spatial metric tensor. $\zeta$ and $\eta$ are the bulk and shear viscosity coefficients, which depend on the equation of state (cf. for instance, \refcite{CutlerEtAl1990}). Bulk viscosity is a result of the fluid trying to restore $\beta$-equilibrium and operates at high temperatures, as opposed to shear viscosity, which is due to particle scattering and is dominant at low temperatures.

For normal nuclear matter, comprising (nonsuperfluid) neutrons, (nonsuperconducting) protons, and electrons, neutron collisions make the biggest contribution to shear viscosity, and the two coefficients are given by \cite{IpserLindblom1991,Sawyer1989,FlowersItoh1979}
\begin{equation}
	\zeta=6\times 10^{-59}\,\rho^2\omega^{-2}T^6\;\mathrm{g}\,\mathrm{cm}^{-1}\,\mathrm{s}^{-1} \label{zeta}
\end{equation} 
and
\begin{equation}
	\eta=347\,\rho^{9/4}T^{-2}\;\mathrm{g}\,\mathrm{cm}^{-1}\,\mathrm{s}^{-1}, \label{eta}
\end{equation} 
where $T$ is the stellar temperature and all the quantities have cgs units. For superfluid nuclear matter another dissipation mechanism dominates, called \emph{mutual friction}. This is expected to occur for temperatures $\lesssim 10^9\,\mathrm{K}$ and suppresses the instability very efficiently \cite{LindblomMendell1995}. Here, we only consider normal nuclear matter; as shown by \refcite{PassamontiEtAl2013}, the star may never enter the superfluid region, since neutrino cooling is balanced by the oscillation-induced viscous heating before the star reaches the transition temperature.\footnote{Reference \cite{PassamontiEtAl2013} uses $E=10^{-4}\,M\Omega^2 R^2$ for the saturation energy of the $f$-mode. However, if the saturation energy is smaller, viscous heating due to the oscillation balances neutrino cooling at lower temperatures.}

\begin{figure}[t]
	\includegraphics[width=0.48\textwidth,keepaspectratio]{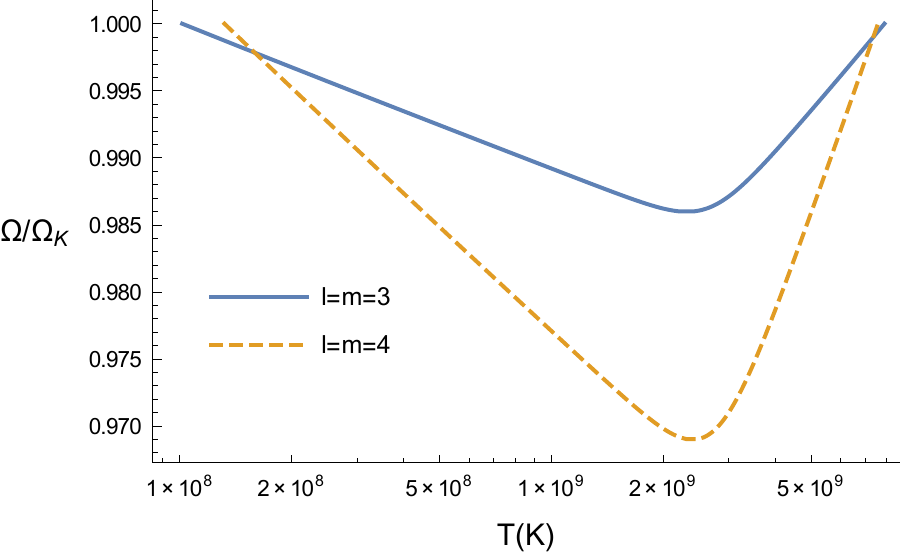}
	\caption{Instability windows of the $l=m=3$ and $l=m=4$ $f$-modes, for a polytropic model with $\Gamma=2$ and $\Gamma_1=2.1$ (the $l=m=2$ $f$-mode does not become unstable for this model). Fiducial values were used for the mass and radius of the star, i.e. $M=1.4\,M_\odot$ and $R=10\,\mathrm{km}$. The angular velocity is normalized to the Kepler limit $\Omega_\mathrm{K}$. These curves were \emph{not} produced using the slow-rotation formalism described in \cref{subsec:The slow-rotation approximation}, because the modes fail to become unstable in this approximation. Also, although this model does not favor the instability, making use of realistic equations of state and relativity can push the windows to quite lower angular velocity values \cite{GaertigEtAl2011,DonevaEtAl2013}.}
	\label{fig:model 7 windows}
\end{figure}

The instability is active only if the total energy rate of the mode is positive, i.e.
\begin{equation}
	\der{E}{t}=\left(\der{E}{t}\right)_\mathrm{GW}+\left(\der{E}{t}\right)_\mathrm{BV}+\left(\der{E}{t}\right)_\mathrm{SV}>0. \label{instability condition}
\end{equation}
By solving this inequality, one obtains the instability window of the mode, namely the region in the $T$-$\Omega$ plane where the mode is CFS unstable (\cref{fig:model 7 windows}). Once the star enters this area, the amplitude of the mode will grow, until such a point where nonlinear effects become important and saturate it. This will be discussed in the following section.


\section{Mode coupling---Quadratic perturbation scheme} \label{sec:Mode coupling---Quadratic perturbation scheme}

Considering the perturbations as small, the modes of the star are uncoupled oscillations (in the nonrotating limit). This is a result of the linear approximation used to define them (cf. \cref{sec:The oscillation modes---Linear perturbation scheme}). However, as the amplitude of the unstable mode grows, the linear approximation fails to accurately describe it; higher-order terms are bound to play an important role in the amplitude evolution, since they introduce mode coupling. The result of this interaction of the unstable mode with other modes is the eventual saturation of the unstable mode's amplitude.

The actual value of this saturation amplitude is mainly important for two reasons. First, it sets the maximum amplitude of the gravitational wave signal obtained from the unstable mode. Second, it affects the evolutionary path of the neutron star inside the instability window. After the star enters the instability window, it cools down, until neutrino cooling is balanced by viscous heating due to the oscillation. Then, it descends the instability window at almost constant temperature, by losing angular momentum. However, magnetic braking also slows down the star, competing with gravitational radiation; as shown in \refcite{PassamontiEtAl2013}, the instability may not have enough time to grow, if the spin-down of the star is dominated by the magnetic torque.

As in previous work for the $r$-mode instability \cite{SchenkEtAl2001,Morsink2002,ArrasEtAl2003,BrinkEtAl2004,BrinkEtAl2004b,BrinkEtAl2005,BondarescuEtAl2007,BondarescuEtAl2009}, we will consider quadratic perturbations and study their effects in the evolution of the $f$-mode. Even higher than second-order terms could, in principle, be important at large oscillation amplitudes, but the complexity of the formulation and the requirements of our problem allow us to choose simplicity over accuracy. Work that also includes cubic nonlinearities can be found in \refcite{BuchlerGoupil1984,vanHoolstSmeyers1993,*vanHoolst1994,*vanHoolst1994b}. Also, for a more general investigation of systems with quadratic and cubic nonlinearities the reader is referred to Chapter 6 of \refcite{NayfehMook1979}.


\subsection{Mode decomposition} \label{subsec:Mode decomposition}

As mentioned in \cref{subsec:The nonrotating limit}, operator $\bm{\mathcal{C}}$ of \cref{eigenvalue equation} is Hermitian. This means that, in the nonrotating limit (where $\bm{\mathcal{B}}$ vanishes), any perturbation, described by the displacement vector $\bm{\xi}(\bm{r},t)$, can be decomposed as
\begin{equation}
	\bm{\xi}(\bm{r},t)=\sum_\alpha q_\alpha(t)\bm{\xi}_\alpha(\bm{r})e^{i\omega_\alpha t}, \label{mode decomposition in the non-rotating limit}
\end{equation}
where $\bm{\xi}_\alpha(\bm{r})$ is a solution to \cref{eigenvalue equation} (with vanishing $\bm{\mathcal{B}}$) and represents the eigenfunction of an oscillation mode, whereas $q_\alpha(t)$ is the \emph{amplitude coefficient}. In the case of polar modes, this eigenfunction is given by \cref{polar mode eigenfunction}.

If rotation is included, operator $\bm{\mathcal{B}}$ is nonvanishing and the solutions to \cref{eigenvalue equation} are not orthogonal, in general. However, instead of a configuration space mode expansion, like \cref{mode decomposition in the non-rotating limit}, one can use a phase space mode expansion \cite{DysonSchutz1979}. Then, a perturbation can be decomposed as \cite{SchenkEtAl2001}
\begin{align}
	\left[
	\begin{array}{c}
		\bm{\xi}(\bm{r},t) \\
		\dot{\bm{\xi}}(\bm{r},t)
	\end{array}
	\right]=\sum_\alpha & \left\{ Q_\alpha(t)\left[
	\begin{array}{c}
		\bm{\xi}_\alpha(\bm{r}) \\
		i\omega_\alpha\bm{\xi}_\alpha(\bm{r})
	\end{array}
	\right]e^{i\omega_\alpha t} \right. \notag \\
	+ \; & \left. Q_\alpha^*(t)\left[
	\begin{array}{c}
		\bm{\xi}_\alpha^*(\bm{r}) \\
		-i\omega_\alpha\bm{\xi}_\alpha^*(\bm{r})
	\end{array}
	\right]e^{-i\omega_\alpha t} \right\}. \label{mode decomposition}
\end{align}
This result was obtained by using the fact that both $(\omega_\alpha,\bm{\xi}_\alpha)$ and $(-\omega_\alpha,\bm{\xi}^*_\alpha)$ are solutions to \cref{eigenvalue equation}, as well as assuming that $\bm{\xi}(\bm{r},t)$ is real.


\subsection{Equations of motion} \label{subsec:Equations of motion}

Including second-order perturbative terms in \cref{equation of motion}, one obtains the quadratic equation of motion, which can be generally written as
\begin{equation}
	\ddot{\bm{\xi}}+\bm{\mathcal{B}}(\dot{\bm{\xi}})+\bm{\mathcal{C}}(\bm{\xi})+\bm{\mathcal{N}}=\bm{0}, \label{quadratic equation of motion}
\end{equation}
where $\bm{\mathcal{N}}$ collectively denotes all $\mathcal{O}(\bm{\xi}^2)$ terms. Substituting \cref{mode decomposition}, and using the eigenvalue equation \eqref{eigenvalue equation} and the orthogonality condition \eqref{mode orthogonality condition}, we get
\begin{equation}
	\dot{Q}_\alpha(t)=\frac{i}{b_\alpha}\langle\bm{\xi}_\alpha,\bm{\mathcal{N}}\rangle e^{-i\omega_\alpha t}. \label{general amplitude equation of motion}
\end{equation}
This is the equation of motion for the amplitude of the mode $Q_\alpha$. If quadratic terms are ignored (or, equivalently, if the perturbation is small), then the amplitude $Q_\alpha$ is constant, since there is no interaction with other modes. However, a nonzero $\bm{\mathcal{N}}$ couples the mode denoted by $\alpha$ with other modes, leading to an energy exchange between them. For a derivation of the equations of motion \eqref{quadratic equation of motion} and \eqref{general amplitude equation of motion}, cf. \cref{sec:Derivation of the equations of motion}.

By further replacing \cref{mode decomposition} in $\bm{\mathcal{N}}$, we obtain
\begin{align}
	\dot{Q}_\alpha(t)=\frac{i}{b_\alpha}\sum_\beta\sum_\gamma\Big[\mathcal{F}_{\alpha\beta\gamma}Q_\beta Q_\gamma e^{i(-\omega_\alpha+\omega_\beta+\omega_\gamma)t} & \notag \\ +\mathcal{F}_{\alpha\bar{\beta}\gamma}Q_\beta^* Q_\gamma e^{i(-\omega_\alpha-\omega_\beta+\omega_\gamma)t} & \notag \\
	+\mathcal{F}_{\alpha\beta\bar{\gamma}}Q_\beta Q_\gamma^* e^{i(-\omega_\alpha+\omega_\beta-\omega_\gamma)t} & \notag \\
	+\mathcal{F}_{\alpha\bar{\beta}\bar{\gamma}}Q_\beta^* Q_\gamma^* e^{i(-\omega_\alpha-\omega_\beta-\omega_\gamma)t} & \Big], \label{amplitude equation of motion with all terms}
\end{align}
where $\mathcal{F}$ denotes the \emph{coupling coefficient} and is generally given by
\begin{equation}
	\mathcal{F}_{\alpha\beta\gamma}=\langle\bm{\xi}_\alpha,\bm{\mathcal{N}}(\bm{\xi}_\beta,\bm{\xi}_\gamma) \rangle. \label{general coupling coefficient}
\end{equation} 
Borrowing the notation of \refcite{SchenkEtAl2001}, a bar over an index means that the corresponding mode eigenfunction in $\bm{\mathcal{N}}$ has to be complex conjugated and its frequency sign reversed. The explicit form of the coupling coefficient is given in \cref{sec:The coupling coefficient}.

Observing \cref{amplitude equation of motion with all terms}, we see that modes couple in triplets, which is a natural consequence of the quadratic-perturbation approximation. This does not, however, restrict the number of couplings for a single mode; if a mode couples to a pair of other modes, it can simultaneously couple to other pairs as well. Also, one can notice that not all terms of \cref{amplitude equation of motion with all terms} are equally significant. Rapidly varying terms do not contribute much on long-term dynamics and average to zero, as opposed to slowly oscillating components (this is proven by means of the multiscale method in \cref{subsec:The multiscale method}). Hence, couplings which really affect the mode amplitude evolution ought to satisfy a resonance condition, e.g.
\begin{equation}
	\omega_\alpha=\omega_\beta+\omega_\gamma+\Delta\omega, \label{resonance condition}
\end{equation} 
where $\Delta\omega$ is a small detuning ($\Delta\omega\ll\omega_i$). Assuming such a relation between the mode frequencies, we can single out a mode triplet and follow its evolution. The amplitude equations of motion for the three modes are
\begin{subequations}
	\label[subequations]{original equations of motion}
	\begin{align}
		\dot{Q}_\alpha & =\frac{i\mathcal{F}_{\alpha\beta\gamma}}{b_\alpha}Q_\beta Q_\gamma e^{-i\Delta\omega t}, \label{mode alpha original equation of motion} \\
		\dot{Q}_\beta & =\frac{i\mathcal{F}_{\beta\bar{\gamma}\alpha}}{b_\beta} Q_\gamma^* Q_\alpha e^{i\Delta\omega t}, \label{mode beta original equation of motion} \\
		\dot{Q}_\gamma & =\frac{i\mathcal{F}_{\gamma\alpha\bar{\beta}}}{b_\gamma} Q_\alpha Q_\beta^* e^{i\Delta\omega t}. \label{mode gamma original equation of motion}
	\end{align}
\end{subequations}

So far, we have assumed that the modes are simply harmonic oscillations, unaffected by any growth/damping mechanisms. However, as discussed in the previous section, all the modes are influenced by various effects, such as gravitational radiation and viscosity. The majority of the modes is damped by these mechanisms, whereas a handful of modes can become unstable and grow, for a certain parameter range.

Such effects are often parametrized by the imaginary part of the oscillation frequency. But we have hitherto assumed that mode frequencies are real, since no such effects were introduced in our equations. So, in order to calculate growth/damping rates, we will use the definition of the corotating-frame mode energy, which is given by \cite{SchenkEtAl2001}
\begin{align}
	E_\alpha & =|Q_\alpha|^2\omega_\alpha b_\alpha \notag \\
	& =|Q_\alpha|^2\omega_\alpha\left[2\omega_\alpha\langle\bm{\xi}_\alpha,\bm{\xi}_\alpha\rangle-\langle\bm{\xi}_\alpha,i\bm{\mathcal{B}}(\bm{\xi}_\alpha)\rangle\right]. \label{mode energy}
\end{align}
This is a quadratic functional of $\bm{\xi}$, so, if $\gamma$ is the imaginary part of the frequency, then
\begin{equation}
	\der{E_\alpha}{t}=2\gamma_\alpha E_\alpha. \label{definition of growth/damping rates}
\end{equation}
Formulas for $\mathrm{d}E/\mathrm{d}t$ for the various mechanisms were provided in the previous section, so we can calculate the growth/damping rate $\gamma$ for a particular mode.

Incorporating the growth/damping rates in \cref{original equations of motion}, we get
\begin{subequations}
	\label[subequations]{equations of motion}
	\begin{align}
		\dot{Q}_\alpha & =\gamma_\alpha Q_\alpha+\frac{i\mathcal{H}}{b_\alpha}Q_\beta Q_\gamma e^{-i\Delta\omega t}, \label{mode alpha equation of motion} \\
		\dot{Q}_\beta & =\gamma_\beta Q_\beta+\frac{i\mathcal{H}}{b_\beta} Q_\gamma^* Q_\alpha e^{i\Delta\omega t}, \label{mode beta equation of motion} \\
		\dot{Q}_\gamma & =\gamma_\gamma Q_\gamma+\frac{i\mathcal{H}}{b_\gamma} Q_\alpha Q_\beta^* e^{i\Delta\omega t}, \label{mode gamma equation of motion}
	\end{align}
\end{subequations}
where we also replaced the coupling coefficients with $\mathcal{H}\equiv\mathcal{F}_{\alpha\beta\gamma}=\mathcal{F}_{\beta\bar{\gamma}\alpha}=\mathcal{F}_{\gamma\alpha\bar{\beta}}$ (cf. \cref{sec:The coupling coefficient}).

Such three-mode systems can give an estimate of the effects of nonlinear coupling to the amplitude of an unstable mode, like the $f$-mode. Such a mode, which we shall call ``parent'', has $\gamma>0$ and has to be coupled to two ``daughter'' modes, which are linearly damped ($\gamma<0$). The efficiency of the coupling depends on the value of the coupling coefficient $\mathcal{H}$, as well as on how close to resonance the three modes are. As we will see, some additional conditions have to be met, in order for the triplet to reach an equilibrium and saturate.


\subsection{Mode normalization} \label{subsec:Mode normalisation}

For the amplitude coefficients of the modes $Q$ to be meaningful, we first have to normalize all the modes according to some convention. By doing this, we will be able to compare the modes, using the same standards. The most popular normalization choice is to fix the mode energy \eqref{mode energy} at unit amplitude to some arbitrary value $E_\mathrm{unit}$, namely,
\begin{equation}
	\omega_\alpha b_\alpha=E_\mathrm{unit}, \label{normalisation}
\end{equation}
for all modes. \Refcite{Morsink2002,ArrasEtAl2003,BrinkEtAl2004,BrinkEtAl2004b,BrinkEtAl2005} use $E_\mathrm{unit}=M\Omega^2 R^2$, whereas \refcite{PassamontiEtAl2013} also uses $E_\mathrm{unit}=Mc^2$. The conversion between two different normalization choices can be straightforwardly written as
\begin{equation}
	|Q_\alpha|^2 E_\mathrm{unit}=|Q'_\alpha|^2 E'_\mathrm{unit}. \label{normalisation conversion}
\end{equation}

Using a normalization choice of the form \eqref{normalisation}, we can rewrite \cref{equations of motion} as
\begin{subequations}
	\label[subequations]{equations of motion with normalisation choice}
	\begin{align}
		\dot{Q}_\alpha & =\gamma_\alpha Q_\alpha+\frac{i\omega_\alpha\mathcal{H}}{E_\mathrm{unit}}Q_\beta Q_\gamma e^{-i\Delta\omega t}, \label{mode alpha equation of motion with normalisation choice} \\
		\dot{Q}_\beta & =\gamma_\beta Q_\beta+\frac{i\omega_\beta\mathcal{H}}{E_\mathrm{unit}} Q_\gamma^* Q_\alpha e^{i\Delta\omega t}, \label{mode beta equation of motion with normalisation choice} \\
		\dot{Q}_\gamma & =\gamma_\gamma Q_\gamma+\frac{i\omega_\gamma\mathcal{H}}{E_\mathrm{unit}} Q_\alpha Q_\beta^* e^{i\Delta\omega t}. \label{mode gamma equation of motion with normalisation choice}
	\end{align}
\end{subequations}
From this form of the amplitude equations of motion it is easier to see that the coupling coefficient $\mathcal{H}$ has units of energy. For the sake of generalization, though, we will be using \cref{equations of motion} in the subsequent sections.\footnote{If one chooses a normalization of the form \eqref{normalisation}, they can simply replace $\mathcal{H}/b_\alpha$ with $\omega_\alpha\mathcal{H}/E_\mathrm{unit}$ in the following sections.}


\subsection{Coupling selection rules} \label{subsec:Coupling selection rules}

As we already mentioned, the three modes forming the coupled network have to obey a resonance condition, given by \cref{resonance condition}. The structure of the coupling coefficient imposes two more conditions, which have to be met in order for the coupling to occur.

As shown in \cref{sec:The coupling coefficient}, the angular dependence of the zeroth-order component of the coupling coefficient has the form
\begin{equation*}
	\iint Y_{l_\alpha}^{*m_\alpha}Y_{l_\beta}^{m_\beta}Y_{l_\gamma}^{m_\gamma}\sin\theta\mathrm{d}\theta\mathrm{d}\phi,
\end{equation*}
where $Y_l^m$ is the spherical-harmonic angular dependence of each mode [cf. \cref{polar mode eigenfunction}]. This integral is proportional to the Clebsch-Gordan coefficients (cf. for instance, \refcite{AbramowitzStegun1972}) and is nonzero if
\begin{equation}
	m_\alpha=m_\beta+m_\gamma \label{m selection rule}
\end{equation} 
and
\begin{equation}
	l_i=l_j+l_k-2\lambda, \label{l selection rule}
\end{equation}
where
\begin{equation*}
	l_i\ge l_j\ge l_k \quad \textrm{and} \quad \lambda=0,1,\ldots\lambda_\mathrm{max}\le\frac{l_k}{2}.
\end{equation*}
\Cref{resonance condition,m selection rule,l selection rule} constitute the selection rules which the coupled mode triplet has to satisfy and restrict the search for possible couplings.\footnote{It should be noted that, even though we evaluated the coupling coefficient in the nonrotating limit in \cref{sec:The coupling coefficient}, these selection rules are valid to all orders in $\Omega$, as shown by \refcite{SchenkEtAl2001}.}


\subsection{Parametric resonance instability} \label{subsec:Parametric resonance instability}

As mentioned before, we are particularly interested in the case where an unstable parent mode ($\gamma_\alpha>0$) is coupled to two damped daughter modes ($\gamma_{\beta,\gamma}<0$). In the beginning of the evolution, when the amplitudes are small, linear terms dominate: the amplitude of the parent grows and the amplitudes of the daughters decrease. At some point, nonlinear terms catch up and the parent starts pumping energy into the daughters. This point occurs when the parent exceeds a certain amplitude, called the \emph{parametric instability threshold}. Such an interaction between the modes is an example of a \emph{parametric resonance instability}, i.e. an instability which can occur when the parameters of an oscillator vary in time (cf. for example, \refcite{LandauLifshitz1969}).

In order to obtain the parametric instability threshold, we take the daughters' equations of motion \eqref{mode beta equation of motion} and \eqref{mode gamma equation of motion} and ask what the value of the parent's amplitude $Q_\alpha$ should be, in order for the daughters' amplitudes $Q_{\beta,\gamma}$ to start growing. Setting $Q_{\beta,\gamma}=\widetilde{Q}_{\beta,\gamma}\exp(i\Delta\omega t/2)$ and writing these equations in matrix form, we get \cite{Dziembowski1982}
\begingroup
\begin{equation*}
	\left(
	\renewcommand*{\arraystretch}{1.4}
	\begin{array}{c}
		\dot{\widetilde{Q}}_\beta \\
		\dot{\widetilde{Q}}\phantom{)}\!\!_\gamma^*
	\end{array}
	\right)=\left(
	\renewcommand*{\arraystretch}{1.5}
	\begin{array}{cc}
		\gamma_\beta-i\Delta\omega/2 & iQ_\alpha\mathcal{H}/b_\beta \\
		 -iQ_\alpha^*\mathcal{H}/b_\gamma & \gamma_\gamma+i\Delta\omega/2
	\end{array}
	\right)\left(
	\renewcommand*{\arraystretch}{1.5}
	\begin{array}{c}
		\widetilde{Q}_\beta \\
		\widetilde{Q}_\gamma^*
	\end{array}
	\right)
\end{equation*}
\endgroup
($Q_\alpha$ is considered an unknown constant). The eigenvalues of the system matrix are
\begin{equation*}
	\lambda_{1,2}=\frac{1}{2}\left[\gamma_\beta+\gamma_\gamma\pm\sqrt{\left(\gamma_\gamma-\gamma_\beta+i\Delta\omega\right)^2+\frac{4\mathcal{H}^2}{b_\beta b_\gamma}|Q_\alpha|^2}\;\right].
\end{equation*}
Then, for the system to admit a growing exponential solution, i.e. for the daughter modes to grow, the condition $\mathrm{Re}(\lambda)>0$ has to be satisfied, for at least one of the eigenvalues. This gives
\begin{equation}
	|Q_\alpha|^2>\frac{\gamma_\beta \gamma_\gamma b_\beta b_\gamma}{\mathcal{H}^2}\left[1+\left(\frac{\Delta\omega}{\gamma_\beta+\gamma_\gamma}\right)^2\right], \label{PIT}
\end{equation}
which is the expression for the parametric instability threshold (PIT), i.e. the amplitude that the parent has to surpass so that the daughters will start growing.\footnote{Note the importance of the mode frequency signs here: if $\omega_\beta\omega_\gamma<0$, then $b_\beta b_\gamma<0$ and no parametric instability can occur. This is a result of the assumed resonance \eqref{resonance condition} between the parent and the daughters. If we perform the same analysis, for example, for mode $\beta$ being the parent, then $\omega_\beta\approx\omega_\alpha-\omega_\gamma$, in which case $\omega_\alpha\omega_\gamma<0$ is a \emph{necessary} condition for parametric instability.}

Ignoring nonlinear effects until the PIT-crossing, parent growth is described by $\dot{Q}_\alpha=\gamma_\alpha Q_\alpha$, which means that PIT-crossing occurs at
\begin{equation}
	t_\mathrm{PIT}=\frac{1}{\gamma_\alpha}\ln\left[\frac{Q_\mathrm{PIT}}{Q_\alpha(0)}\right], \label{PIT-crossing time}
\end{equation}
where $Q_\alpha(0)$ is the parent's initial amplitude.


\subsection{Equilibrium solution} \label{subsec:Equilibrium solution}

Once the parent crosses the PIT and the daughters start growing, the three modes will continue interacting by exchanging energy. There can be two general outcomes from this process: (i) the system admits a stable equilibrium solution and all three modes reach saturation, or (ii) the parent's growth cannot be halted by the daughters and all three modes grow, continuing to exchange energy.

\Cref{equations of motion} admit an easy-to-obtain equilibrium solution. Expressing the complex amplitudes $Q$ in terms of real amplitude and phase variables, we can introduce the variable transformation \cite{Dziembowski1982}
\begin{subequations}
	\label[subequations]{amplitude and phase variables}
	\begin{align}
		Q_\alpha=\frac{\sqrt{b_\beta b_\gamma}}{\mathcal{H}}\varepsilon_\alpha e^{i\vartheta_\alpha}, \label{mode alpha amplitude and phase variables} \\
		Q_\beta=\frac{\sqrt{b_\gamma b_\alpha}}{\mathcal{H}}\varepsilon_\beta e^{i\vartheta_\beta}, \label{mode beta amplitude and phase variables} \\
		Q_\gamma=\frac{\sqrt{b_\alpha b_\beta}}{\mathcal{H}}\varepsilon_\gamma e^{i\vartheta_\gamma}. \label{mode gamma amplitude and phase variables}
	\end{align}
\end{subequations}
Then, \cref{equations of motion} are written as
\begin{subequations}
	\label[subequations]{equations of motion with amplitude and phase variables}
	\begin{align}
		\dot{\varepsilon}_\alpha=\gamma_\alpha\varepsilon_\alpha+\varepsilon_\beta\varepsilon_\gamma\sin\varphi, \label{mode alpha equation of motion with amplitude and phase variables} \\
		\dot{\varepsilon}_\beta=\gamma_\beta\varepsilon_\beta-\varepsilon_\gamma\varepsilon_\alpha\sin\varphi, \label{mode beta equation of motion with amplitude and phase variables} \\
		\dot{\varepsilon}_\gamma=\gamma_\gamma\varepsilon_\gamma-\varepsilon_\alpha\varepsilon_\beta\sin\varphi, \label{mode gamma equation of motion with amplitude and phase variables}
	\end{align}
	and
	\begin{equation}
		\dot{\varphi}=\cot\varphi\left[\frac{\dot{\varepsilon}_\alpha}{\varepsilon_\alpha}+\frac{\dot{\varepsilon}_\beta}{\varepsilon_\beta}+\frac{\dot{\varepsilon}_\gamma}{\varepsilon_\gamma}-\gamma\right]+\Delta\omega, \label{phase equation of motion}
	\end{equation}
\end{subequations}
where $\varphi=\vartheta_\alpha-\vartheta_\beta-\vartheta_\gamma+\Delta\omega t$ and $\gamma=\gamma_\alpha+\gamma_\beta+\gamma_\gamma$. Setting the time derivatives to zero, we find the steady-state solution
\begin{subequations}
	\label[subequations]{amplitude and phase variables equilibria}
	\begin{align}
		\varepsilon_\alpha^2 & =\gamma_\beta\gamma_\gamma\left[1+\left(\frac{\Delta\omega}{\gamma}\right)^2\right], \label{mode alpha real amplitude equilibrium} \\
		\varepsilon_\beta^2 & =-\gamma_\gamma\gamma_\alpha\left[1+\left(\frac{\Delta\omega}{\gamma}\right)^2\right], \label{mode beta real amplitude equilibrium} \\
		\varepsilon_\gamma^2 & =-\gamma_\alpha\gamma_\beta\left[1+\left(\frac{\Delta\omega}{\gamma}\right)^2\right], \label{mode gamma real amplitude equilibrium}
	\end{align}
	and
	\begin{equation}
		\cot\varphi=\frac{\Delta\omega}{\gamma}, \label{phase equilibium}
	\end{equation} 
\end{subequations}
or, in terms of the original complex amplitudes,
\begin{subequations}
	\label[subequations]{equilibrium amplitudes}
	\begin{align}
		|Q_\alpha|^2 & =\frac{\gamma_\beta \gamma_\gamma b_\beta b_\gamma}{\mathcal{H}^2}\left[1+\left(\frac{\Delta\omega}{\gamma}\right)^2\right], \label{mode alpha equilibrium amplitude} \\
		|Q_\beta|^2 & =-\frac{\gamma_\gamma \gamma_\alpha b_\gamma b_\alpha}{\mathcal{H}^2}\left[1+\left(\frac{\Delta\omega}{\gamma}\right)^2\right], \label{mode beta equilibrium amplitude} \\
		|Q_\gamma|^2 & =-\frac{\gamma_\alpha \gamma_\beta b_\alpha b_\beta}{\mathcal{H}^2}\left[1+\left(\frac{\Delta\omega}{\gamma}\right)^2\right]. \label{mode gamma equilibrium amplitude}
	\end{align}
\end{subequations}
Note that, for $|\gamma_\beta+\gamma_\gamma|\gg\gamma_\alpha$, the equilibrium amplitude \eqref{mode alpha equilibrium amplitude} of the unstable mode coincides with the PIT \eqref{PIT}.


\subsection{Saturation conditions} \label{subsec:Saturation conditions}

Such three-mode coupled systems, exhibiting a parametric resonance instability, have been studied in the past \cite{WersingerEtAl1980,*WersingerEtAl1980b,Dimant2000} for their significance in various fields, e.g. plasma physics \cite{Anderson1976,Verheest1976}. These studies show that certain conditions have to be met, in order for the system to approach saturation.

\begin{figure*}[t]
	\includegraphics[width=\textwidth,keepaspectratio]{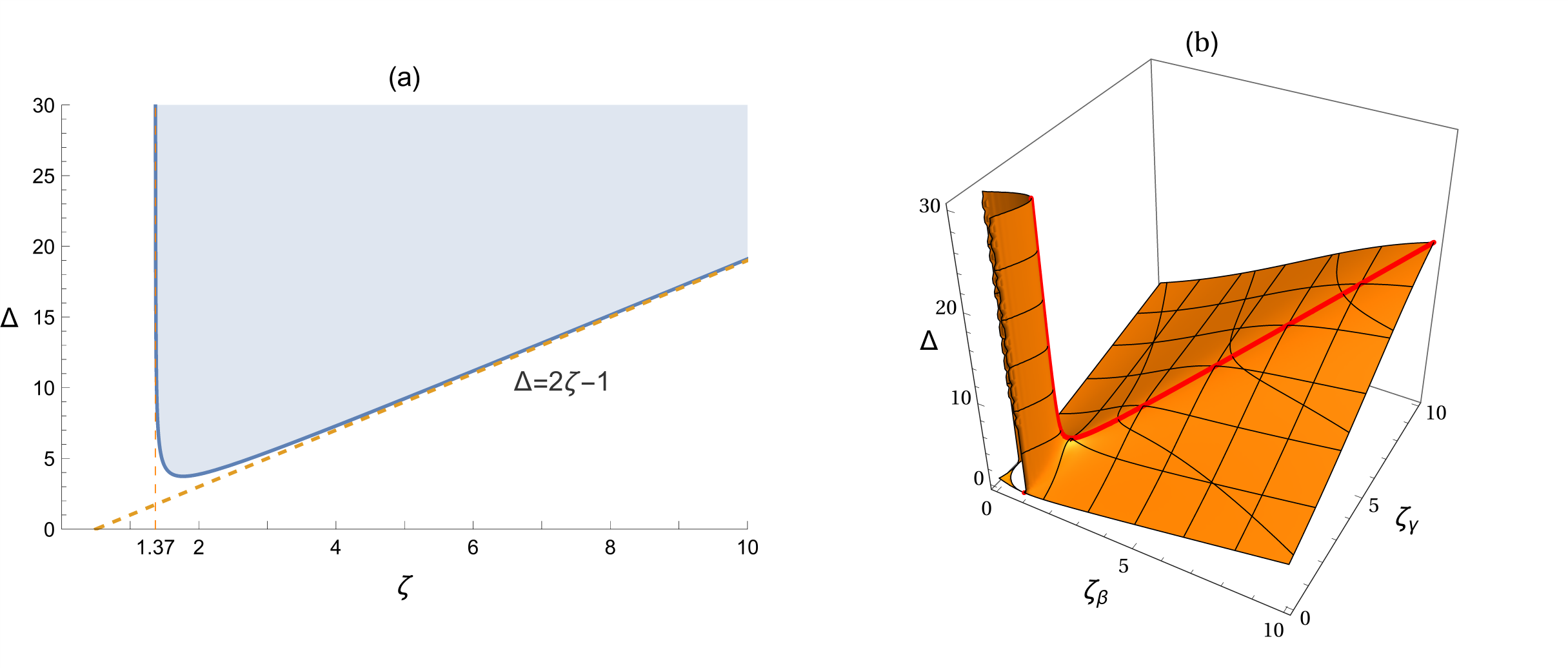}
	\caption{(a) $\varDelta$ versus $\zeta\,(\equiv\zeta_\beta=\zeta_\gamma)$. The saturation condition \eqref{saturation condition 2.2} is satisfied inside the shaded area. The two asymptotes at $\zeta\approx 1.37$ and $\varDelta=2\zeta-1$ are also shown (dashed lines). A global minimum occurs at $(1.77,3.73)$. (b) $\varDelta$ versus $\zeta_\beta$ versus $\zeta_\gamma$. The saturation condition \eqref{saturation condition 2} is satisfied inside the region that lies above the plotted surface. The thick line corresponds to the case where $\zeta_\beta=\zeta_\gamma\equiv\zeta$.}
	\label{fig:saturation condition}
\end{figure*}

Performing a linear stability analysis of \cref{equations of motion with amplitude and phase variables} (which is presented in \cref{subsec:Linear stability analysis}), we find that the equilibrium solution \eqref{equilibrium amplitudes} is stable if \cite{Dziembowski1982}
\begin{equation}
	|\gamma_\beta+\gamma_\gamma|>\gamma_\alpha \label{saturation condition 1}
\end{equation}
and
\begin{widetext}
	\begin{align}
		3 & \left\{\left(\zeta_\beta+\zeta_\gamma-1\right)\left[\left(\zeta_\beta-\zeta_\gamma\right)^2+2\left(\zeta_\beta+\zeta_\gamma\right)+1\right]-6\zeta_\beta\zeta_\gamma\right\}\left(\frac{\Delta\omega}{\gamma}\right)^4 \notag \\
		+ & \left\{\left(\zeta_\beta+\zeta_\gamma-1\right)\left[\left(\zeta_\beta-\zeta_\gamma\right)^2+\left(\zeta_\beta+\zeta_\gamma\right)^2+2\right]-12\zeta_\beta\zeta_\gamma\right\}\left(\frac{\Delta\omega}{\gamma}\right)^2 \notag \\
		- & \left(\zeta_\beta+\zeta_\gamma-1\right)^3-2\zeta_\beta\zeta_\gamma>0, \label{saturation condition 2}
	\end{align} 
\end{widetext}
where $\zeta_{\beta,\gamma}=-\gamma_{\beta,\gamma}/\gamma_\alpha$, which are the relative damping rates of the daughters. To simplify the expression above, we set $\zeta\equiv\zeta_\beta=\zeta_\gamma$. Then, keeping in mind that \cref{saturation condition 1} should also be true, it is reduced to
\begin{equation}
	\zeta>\frac{1+\sqrt{3}}{2}\approx 1.37 \label{saturation condition 2.1}
\end{equation}
and
\begin{equation}
	\varDelta^2>\frac{2\zeta^2-2\zeta+1}{2\zeta^2-2\zeta-1}(1-2\zeta)^2, \label{saturation condition 2.2}
\end{equation}
where $\varDelta=\Delta\omega/\gamma_\alpha$.

First, we notice that \cref{saturation condition 2.1} imposes a stronger constraint on $\zeta$ than \cref{saturation condition 1}. Second, we see from \cref{saturation condition 2.2} that there is a lower limit on the detuning, which depends on $\zeta$. This is illustrated in \cref{fig:saturation condition}.

\begin{figure}[h]
	\includegraphics[width=0.48\textwidth,keepaspectratio]{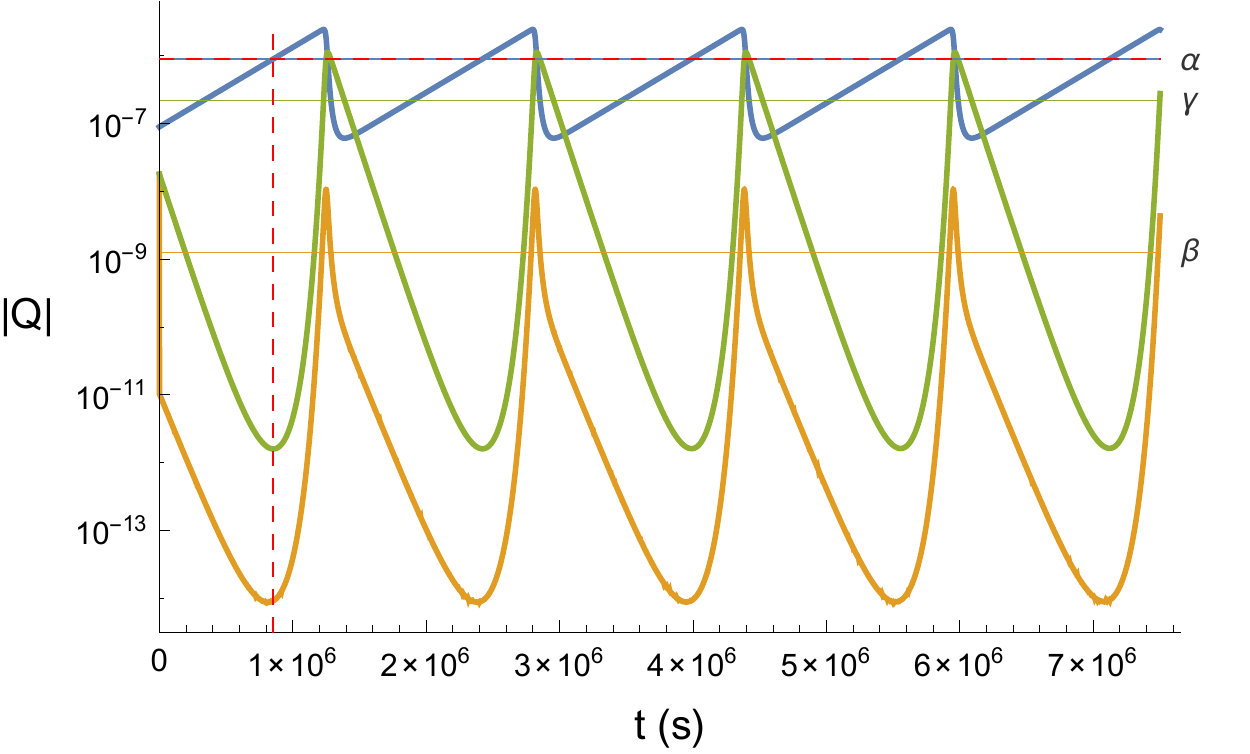}
	\caption{Amplitude evolution of a coupled triplet that satisfies the saturation conditions. Horizontal solid lines represent the saturation amplitudes of each mode. The dashed horizontal line shows the position of the triplet's PIT, whereas the dashed vertical line denotes the PIT-crossing time. At that point the parent (mode $\alpha$) crosses the PIT and the daughters (modes $\beta$ and $\gamma$), which were damped until that point, start to grow. Then, the amplitudes oscillate and finally converge (albeit very slowly in this example) around their equilibrium values (the parent's equilibrium coincides with the PIT in this example). In this graph, we show the triplet with the \emph{lowest PIT}, in a polytropic model with $\Gamma=2$ and $\Gamma_1=2.1$, for $\Omega=\Omega_\mathrm{K}$ and $T=5\times 10^9\,\mathrm{K}$. Mode $\alpha$ is the $\prescript{3}{3}f$-mode, mode $\beta$ is the $\prescript{-4}{4}f$-mode, and mode $\gamma$ is the $\prescript{7}{7}g_5$-mode (where the notation $\prescript{m}{l\,}g_n$ has 
been used). The growth/damping rates are $\gamma_\alpha=2.7\times 10^{-6}\,\mathrm{rad}\,\mathrm{s}^{-1},\,\gamma_\beta=-1.0\,\mathrm{rad}\,\mathrm{s}^{-1}$, and $\gamma_\gamma=-1.4\times 10^{-5}\,\mathrm{rad}\,\mathrm{s}^{-1}$, and the detuning is $\Delta\omega=14.1\,\mathrm{rad}\,\mathrm{s}^{-1}$. The value of $|Q|$ depends on the mode normalization choice as $|Q|=\sqrt{E_\mathrm{mode}/E_\mathrm{unit}}$ (cf. \cref{subsec:Mode normalisation}); here, we chose $E_\mathrm{unit}=Mc^2$.}
	\label{fig:3f-25}
\end{figure}

If \cref{saturation condition 1} is not satisfied, all solutions are unbounded and the triplet's amplitudes grow to infinity; the damping rate of the daughters needs to be larger than the driving rate of the parent, in order to stop its growth. The additional condition \eqref{saturation condition 2} [or, for $\gamma_\beta=\gamma_\gamma$, \eqref{saturation condition 2.1} and \eqref{saturation condition 2.2}] is more unintuitive; as shown by \refcite{WersingerEtAl1980,WersingerEtAl1980b,Dimant2000}, a number of interesting behaviors occur when it is not fulfilled, including limit cycles and chaotic motion. The amplitude evolution of a triplet satisfying the saturation conditions can be seen in \cref{fig:3f-25}.

\section{Discussion} \label{sec:Discussion}


The anticipated advent of gravitational-wave astronomy will hopefully shed some light on the neutron star equation of state problem: should gravitational radiation from individual sources be observable, much information about the neutron star interior could be obtained. However, gravitational-wave asteroseismology would have to deal with very weak signals, generated by stellar oscillations. The fact that some of these oscillations are unstable to the emission of gravitational radiation, due to the CFS mechanism presented in \cref{sec:The f-mode instability}, works to our advantage: the amplitude of the mode will grow until such a point when nonlinear effects saturate the instability.

Studies on the $r$-mode instability have shown that the saturation levels will make detection very difficult. In the most optimistic cases, the signal may be detectable with Advanced LIGO from within the local galaxy group \cite{BondarescuEtAl2007,BondarescuEtAl2009}. As far as the $f$-mode instability is concerned, reasonably high saturation levels make the signal from a nascent star definitely detectable with the Einstein Telescope (in some cases even with Advanced LIGO) for sources in the Virgo cluster \cite{PassamontiEtAl2013}.

Estimating the saturation amplitudes for the $r$- and $f$-mode instabilities is also important for another reason: their values affect the evolution of the star inside the instability area. A newborn star, for which both instabilities can be significant, will enter the instability window, which it will traverse at approximately constant angular velocity, until it reaches thermal equilibrium; then, at approximately constant temperature, the star will spin down due to the emission of gravitational radiation, as well as magnetic braking, until it exits the window. The saturation amplitude affects the duration of these phases, thus the time which the star spends inside the instability area.

By taking quadratic perturbations into account, coupled three-mode networks are formed throughout the star. These triplets have to satisfy an internal resonance and two selection rules for their orders $m$ and degrees $l$. Although any triplet can be part of this network, we are obviously interested in the case where one of the participating modes is the unstable $f$-mode. Then, the coupled triplet is said to be parametrically resonant and can lead to a parametric instability, if the unstable (parent) mode crosses the so-called parametric instability threshold. At that point, the other two (daughter) modes start growing. The system reaches saturation if certain conditions are satisfied for the modes' growth/damping rates, and their frequency mismatch.

In this paper, we have focused on polar modes, like \mbox{$f$-,} \mbox{$p$-,} and $g$-modes. However, all the formulas presented in \cref{sec:Mode coupling---Quadratic perturbation scheme} are also applicable to axial modes. It is only in \cref{sec:The coupling coefficient} where we assume that all three modes are polar, and find an expression for the zeroth-order component of the coupling coefficient. Results from the application of the formulation above to Newtonian, polytropic stars will be presented in a subsequent paper.


\begin{acknowledgments}
	We gratefully acknowledge the support of the German Science Foundation (DFG) via SFB/TR7. We would also like to thank K. Glampedakis for his important comments and suggestions on the manuscript.
\end{acknowledgments}


\appendix

\let\oldsection=\section
\def\section#1{\oldsection{\uppercase{#1}}}               


\section{Derivation of the equations of motion} \label[appsec]{sec:Derivation of the equations of motion}


\subsection{The quadratic equation of motion} \label[appsec]{subsec:The quadratic equation of motion}

The derivation of the quadratic equation of motion \eqref{quadratic equation of motion} can be performed in the same way as the derivation of the linear equation of motion \eqref{equation of motion}, except that now we also want to retain second-order perturbative terms.

Following \refcite{Dziembowski1982}, we will use the velocity $\bm{v}$, instead of the Lagrangian displacement $\bm{\xi}$, to describe the perturbation. As mentioned in \cref{sec:The oscillation modes---Linear perturbation scheme}, the background velocity is zero (because we are working in the corotating frame), so $\bm{v}\equiv\delta\bm{v}=\dot{\bm{\xi}}$. Differentiating \cref{Euler} with respect to time and imposing perturbations on the equilibrium state, we obtain the equation of motion for the velocity, namely
\begin{equation}
	\ddot{\bm{v}}+\bm{\mathcal{B}}(\dot{\bm{v}})+\bm{\mathcal{C}}(\bm{v})+\bm{\mathcal{N}_v}=\bm{0}, \label{equation of motion for the velocity}
\end{equation}
where
\begin{equation}
	\bm{\mathcal{B}}(\bm{v})=2\bm{\Omega}\times\bm{v} \label{operator B with velocity}
\end{equation}
and
\begin{equation}
	\bm{\mathcal{C}}(\bm{v})=\frac{1}{\rho}\nabla\left(\parder{\delta_1 p}{t}\right)-\frac{\nabla p}{\rho^2}\parder{\delta_1\rho}{t}+\nabla\left(\parder{\delta_1\Phi}{t}\right), \label{operator C with velocity}
\end{equation} 
with $\delta_1$ denoting first-order and $\delta_2$ second-order Eulerian perturbations. $\bm{\mathcal{N}_v}$ represents the quadratic terms, which are explicitly written as
\begin{align}
	\bm{\mathcal{N}_v}=\parder{}{t} & \left[(\bm{v}\cdot\nabla)\bm{v}+\frac{\nabla\delta_2 p}{\rho}+\delta_1\left(\frac{1}{\rho}\right)\nabla\delta_1 p \right. \notag \\
	& \left.+\delta_2\left(\frac{1}{\rho}\right)\nabla p+\nabla\delta_2\Phi\right], \label{nonlinear terms}
\end{align}
where
\begin{equation*}
	\delta_1\left(\frac{1}{\rho}\right)=-\frac{\delta_1\rho}{\rho^2}\quad\textrm{and}\quad\delta_2\left(\frac{1}{\rho}\right)=-\frac{\delta_2\rho}{\rho^2}+\frac{(\delta_1\rho)^2}{\rho^3}.
\end{equation*}
It should be noted that $\bm{\mathcal{N}}$, which appears in \cref{quadratic equation of motion}, is related to $\bm{\mathcal{N}_v}$ simply by $\bm{\mathcal{N}_v}=\partial\bm{\mathcal{N}}/\partial t$.

Perturbing the continuity equation \eqref{continuity}, we get
\begin{equation}
	\parder{\delta_1\rho}{t}=-\rho\nabla\cdot\bm{v}-(\bm{v}\cdot\nabla)\rho \label{continuity perturbed first-order}
\end{equation} 
and
\begin{equation}
	\parder{\delta_2\rho}{t}=-\delta_1\rho\nabla\cdot\bm{v}-(\bm{v}\cdot\nabla)\delta_1\rho, \label{continuity perturbed second-order}
\end{equation}
for first- and second-order terms, respectively. Accordingly, the perturbed Poisson equation \eqref{Poisson} gives
\begin{equation*}
	\nabla^2\delta_1\Phi=4\pi G\delta_1\rho \quad\textrm{and}\quad \nabla^2\delta_2\Phi=4\pi G\delta_2\rho,
\end{equation*}
whose (time-differentiated) solutions are
\begin{equation}
	\parder{\delta_1\Phi}{t}=G\int\frac{\nabla_{\bm{r}'}\cdot(\rho\bm{v})}{|\bm{r}-\bm{r}'|}\mathrm{d}^3\bm{r}' \label{Poisson perturbed first-order}
\end{equation} 
and
\begin{equation}
	\parder{\delta_2\Phi}{t}=G\int\frac{\nabla_{\bm{r}'}\cdot(\delta_1\rho\bm{v})}{|\bm{r}-\bm{r}'|}\mathrm{d}^3\bm{r}'. \label{Poisson perturbed second-order}
\end{equation}

Finally, perturbation of the equation of state $p=p(\rho,\mu)$ to second order gives
\begin{equation*}
	\Delta p=\left(\parder{p}{\rho}\right)_\mu\Delta\rho+\frac{1}{2}\left(\nparder{2}{p}{\rho}\right)_\mu(\Delta\rho)^2,
\end{equation*}
or
\begin{equation}
	\frac{\Delta p}{p}=\Gamma_1\frac{\Delta\rho}{\rho}+\frac{1}{2}\left[\Gamma_1(\Gamma_1-1)+\left(\parder{\Gamma_1}{\ln\rho}\right)_\mu\right]\left(\frac{\Delta\rho}{\rho}\right)^2, \label{EoS perturbed to second order}
\end{equation} 
where $\Gamma_1$ is defined by \cref{Gamma_1}. Here we have assumed that $\Delta\mu=0$, i.e. the composition is frozen (if $\mu$ corresponds to the composition) and/or the star is isentropic (if $\mu$ denotes entropy). Also, we have used Lagrangian perturbations, which, to second order, are related to Eulerian by
\begin{equation*}
	\Delta f=\delta_1 f+(\bm{\xi}\cdot\nabla)f+\delta_2 f+(\bm{\xi}\cdot\nabla)\delta_1 f+\frac{1}{2}\bm{\xi}\cdot\left[\bm{\xi}\cdot\nabla\left(\nabla f\right)\right].
\end{equation*}
Using this, we obtain from \cref{EoS perturbed to second order}
\begin{equation}
	\parder{\delta_1 p}{t}=-(\bm{v}\cdot\nabla)p-p\Gamma_1\nabla\cdot\bm{v} \label{EoS perturbed first-order}
\end{equation} 
and
\begin{equation}
	\parder{\delta_2 p}{t}=-(\bm{v}\cdot\nabla)\delta_1 p+\left[(\bm{\xi}\cdot\nabla)(p\Gamma_1)+p\Gamma_1\chi\nabla\cdot\bm{\xi}\right]\nabla\cdot\bm{v}, \label{EoS perturbed second-order}
\end{equation}
where
\begin{equation*}
	\chi=\Gamma_1+\left(\parder{\ln\Gamma_1}{\ln\rho}\right)_\mu.
\end{equation*}


\subsection{The amplitude equation of motion} \label[appsec]{subsec:The amplitude equation of motion}

In order to obtain the equation of motion for the amplitude \eqref{general amplitude equation of motion}, we have to replace $\bm{v}$ in \cref{equation of motion for the velocity} with the expansion \eqref{mode decomposition}. Note that this expansion implies that
\begin{equation}
	\sum_\alpha\left(\dot{Q}_\alpha\bm{\xi}_\alpha e^{i\omega_\alpha t}+\dot{Q}^*_\alpha\bm{\xi}^*_\alpha e^{-i\omega_\alpha t}\right)=\bm{0} \label{mode decomposition condition 1}
\end{equation}
and
\begin{align}
	\sum_\alpha & \left(\ddot{Q}_\alpha\bm{\xi}_\alpha e^{i\omega_\alpha t}+i\omega_\alpha\dot{Q}_\alpha\bm{\xi}_\alpha e^{i\omega_\alpha t}\right. \notag \\
	& \left.+\ddot{Q}^*_\alpha\bm{\xi}^*_\alpha e^{-i\omega_\alpha t}-i\omega_\alpha\dot{Q}^*_\alpha\bm{\xi}^*_\alpha e^{-i\omega_\alpha t}\right)=\bm{0}. \label{mode decomposition condition 2}
\end{align}
Making use of the eigenvalue equation \eqref{eigenvalue equation}, the orthogonality condition \eqref{mode orthogonality condition}, as well as \cref{mode decomposition condition 1,mode decomposition condition 2}, we get
\begin{equation}
	\ddot{Q}_\alpha+i\omega_\alpha\dot{Q}_\alpha=\frac{i}{b_\alpha}\langle\bm{\xi}_\alpha,\bm{\mathcal{N}_v}\rangle e^{-i\omega_\alpha t}. \label{general amplitude equation of motion with velocity}
\end{equation}
It is easily seen that \cref{general amplitude equation of motion with velocity} is obtained by differentiating \cref{general amplitude equation of motion} with respect to time. By further replacing the expansion \eqref{mode decomposition} in $\bm{\mathcal{N}_v}$ one gets
\begin{align}
	\ddot{Q}_\alpha+i\omega_\alpha\dot{Q}_\alpha=\frac{i}{b_\alpha}i\omega_\alpha\sum_{\beta,\gamma}\Big[F_{\alpha\beta\gamma}Q_\beta Q_\gamma e^{i(-\omega_\alpha+\omega_\beta+\omega_\gamma)t} & \notag \\ +F_{\alpha\bar{\beta}\gamma}Q_\beta^* Q_\gamma e^{i(-\omega_\alpha-\omega_\beta+\omega_\gamma)t} & \notag \\
	+F_{\alpha\beta\bar{\gamma}}Q_\beta Q_\gamma^* e^{i(-\omega_\alpha+\omega_\beta-\omega_\gamma)t} & \notag \\
	+F_{\alpha\bar{\beta}\bar{\gamma}}Q_\beta^* Q_\gamma^* e^{i(-\omega_\alpha-\omega_\beta-\omega_\gamma)t} & \Big], \label{amplitude equation of motion with all terms with velocity}
\end{align}
where
\begin{equation}
	F_{\alpha\beta\gamma}=\frac{1}{i\omega_\alpha}\langle\bm{\xi}_\alpha,\bm{\mathcal{N}_v}(\bm{\xi}_\beta,\bm{\xi}_\gamma)\rangle \label{general coupling coefficient with velocity}
\end{equation} 
is the coupling coefficient (a bar over an index means that the corresponding mode eigenfunction in $\bm{\mathcal{N}_v}$ has to be complex conjugated and its frequency sign reversed).

As mentioned in \cref{subsec:Equations of motion}, not all terms in \cref{amplitude equation of motion with all terms with velocity} play an equally important role in the amplitude evolution. As shown in \cref{subsec:The multiscale method}, a resonance condition between the modes is necessary for the dynamics of the system to be significantly affected by quadratic terms. Assuming a resonance of the form $\omega_\alpha=\omega_\beta+\omega_\gamma+\Delta\omega$, where $\Delta\omega$ is a small detuning, one can omit rapidly varying terms in \cref{amplitude equation of motion with all terms with velocity}. Then, choosing a mode triplet which satisfies the resonance condition, we get
\begin{subequations}
	\label[subequations]{original equations of motion with velocity}
	\begin{align}
		\ddot{Q}_\alpha+i\omega_\alpha\dot{Q}_\alpha & =\frac{i}{b_\alpha}i\omega_\alpha F_{\alpha\beta\gamma}Q_\beta Q_\gamma e^{-i\Delta\omega t}, \label{mode alpha original equation of motion with velocity} \\
		\ddot{Q}_\beta+i\omega_\beta\dot{Q}_\beta & =\frac{i}{b_\beta}i\omega_\beta F_{\beta\bar{\gamma}\alpha} Q_\gamma^* Q_\alpha e^{i\Delta\omega t}, \label{mode beta original equation of motion with velocity} \\
		\ddot{Q}_\gamma+i\omega_\gamma\dot{Q}_\gamma & =\frac{i}{b_\gamma}i\omega_\gamma F_{\gamma\alpha\bar{\beta}} Q_\alpha Q_\beta^* e^{i\Delta\omega t}. \label{mode gamma original equation of motion with velocity}
	\end{align}
\end{subequations}
If such a resonance exists, it can be shown that $i\omega_\alpha F_{\alpha\beta\gamma}=i(\omega_\alpha-\Delta\omega)\mathcal{F}_{\alpha\beta\gamma}$, where $\mathcal{F}_{\alpha\beta\gamma}$ is given by \cref{general coupling coefficient}. So, ignoring the detuning, $F_{\alpha\beta\gamma}\approx\mathcal{F}_{\alpha\beta\gamma}$, which also implies that $\ddot{Q}$ is negligible, because only then we can retrieve the equivalent system \eqref{original equations of motion}.

Setting $\mathcal{H}\equiv F_{\alpha\beta\gamma}=F_{\beta\bar{\gamma}\alpha}=F_{\gamma\alpha\bar{\beta}}$ (cf. \cref{sec:The coupling coefficient}) and introducing growth/damping rates for the modes, \cref{original equations of motion with velocity} become
\begin{subequations}
	\label[subequations]{equations of motion with velocity}
	\begin{eqnarray}
		\dot{Q}_\alpha &=& \gamma_\alpha Q_\alpha+\frac{i\mathcal{H}}{b_\alpha}Q_\beta Q_\gamma e^{-i\Delta\omega t}, \label{mode alpha equation of motion with velocity} \\
		\dot{Q}_\beta  &=& \gamma_\beta Q_\beta+\frac{i\mathcal{H}}{b_\beta} Q_\gamma^* Q_\alpha e^{i\Delta\omega t}, \label{mode beta equation of motion with velocity} \\
		\dot{Q}_\gamma &=& \gamma_\gamma Q_\gamma+\frac{i\mathcal{H}}{b_\gamma} Q_\alpha Q_\beta^* e^{i\Delta\omega t}, \label{mode gamma equation of motion with velocity}
	\end{eqnarray}
\end{subequations}
which coincide with \cref{equations of motion}.


\section{The coupling coefficient} \label[appsec]{sec:The coupling coefficient}

Proceeding with the evaluation of \cref{general coupling coefficient with velocity}, using equations from \cref{subsec:The quadratic equation of motion}, we find an explicit form for the coupling coefficient, which is \cite{Dziembowski1982}
\begin{equation}
	F_{\alpha\beta\gamma}=\frac{1}{\omega_\alpha}\left(\omega_\beta S_{\alpha\beta\gamma}+\omega_\gamma S_{\alpha\gamma\beta}\right), \label{mode alpha coupling coefficient}
\end{equation}
where
\begin{widetext}
	\begin{align}
		S_{\alpha\beta\gamma}=\int\Bigg\{ & \rho\omega_\beta\omega_\gamma\big[-\nabla\left(\bm{\xi}_\beta\cdot\bm{\xi}_\gamma\right)+\bm{\xi}_\beta\times\left(\nabla\times\bm{\xi}_\gamma\right)+ \bm{\xi}_\gamma\times\left(\nabla\times\bm{\xi}_\beta\right)\big] \notag \\ 
		& -\frac{1}{\rho}\big[\nabla\cdot\left(\rho\bm{\xi}_\beta\right)\nabla\left(\bm{\xi}_\gamma\cdot\nabla p+p\Gamma_1\nabla\cdot\bm{\xi}_\gamma\right)+\nabla\cdot\left(\rho\bm{\xi}_\gamma\right)\nabla\left(\bm{\xi}_\beta\cdot\nabla p+p\Gamma_1\nabla\cdot\bm{\xi}_\beta\right)\big] \notag \\
		& +\nabla\cdot\left(\rho\bm{\xi}_\beta\right)\nabla\cdot\left(\rho\bm{\xi}_\gamma\right)\frac{\nabla p}{\rho^2}-\left[\bm{\xi}_\beta\cdot\nabla\left(\frac{\nabla\cdot\left(\rho\bm{\xi}_\gamma\right)}{\rho}\right)\right]\nabla p-G\rho\nabla\left[\int\frac{\nabla_{\bm{r}'}\cdot\left[\bm{\xi}_\beta\nabla\cdot\left(\rho\bm{\xi}_\gamma\right)\right]}{|\bm{r}-\bm{r}'|}\mathrm{d}^3\bm{r}'\right] \notag \\
		& +\nabla\big[\bm{\xi}_\beta\cdot\nabla\left(\bm{\xi}_\gamma\cdot\nabla p+p\Gamma_1\nabla\cdot\bm{\xi}_\gamma\right)+\left(\nabla\cdot\bm{\xi}_\beta\right)\bm{\xi}_\gamma\cdot\nabla\left(p\Gamma_1\right) +p\Gamma_1\chi\left(\nabla\cdot\bm{\xi}_\beta\right)\left(\nabla\cdot\bm{\xi}_\gamma\right)\big]\Bigg\}\cdot\bm{\xi}_\alpha^*\mathrm{d}^3\bm{r}. \label{coupling coefficient auxilliary parameter}
	\end{align}
\end{widetext}
The expressions for $F_{\beta\bar{\gamma}\alpha}$ and $F_{\gamma\alpha\bar{\beta}}$ are obtained from \cref{mode alpha coupling coefficient}, keeping in mind that a bar over an index means that the corresponding mode eigenfunction has to be complex conjugated and the corresponding frequency has to change sign.

As pointed out by \refcite{SchenkEtAl2001}, the expression above for the coupling coefficient is identical for both nonrotating and rotating stars. This of course does not make the actual value of the coupling coefficient the same for both cases. If rotation is included, the eigenfrequencies, the eigenfunctions, and the equilibrium quantities are all affected (cf. \cref{subsec:The slow-rotation approximation}).

We now assume that $\bm{\xi}$ takes the form \eqref{polar mode eigenfunction}, namely, it describes the eigenfunction of a polar mode in the nonrotating limit. We also define the dimensionless quantities~\cite{UnnoEtAl1989}
\begin{equation*}
	\begin{array}{ccc}
		x=r/R, & & \tilde{\omega}=\omega/\sqrt{GM/R^3}, \\[.8em]
		\displaystyle y_1=\frac{\xi_r}{r}, & & \displaystyle y_2=c_1\tilde{\omega}^2\frac{\xi_h}{r}, \\[1em]
		\displaystyle y_3=\frac{\delta\Phi}{gr}, & & \displaystyle y_4=\frac{1}{g}\der{\delta\Phi}{r}, \\[1.5em]
		\displaystyle c_1=\left(\frac{r}{R}\right)^3\frac{M}{M_r}, & & \displaystyle U=\der{\ln M_r}{\ln r}=\frac{4\pi\rho r^3}{M_r}, \\[1.5em]
		\displaystyle V_g=\frac{V}{\Gamma_1}=-\frac{1}{\Gamma_1}\der{\ln p}{\ln r}, & & \displaystyle A^*=\frac{1}{\Gamma_1}\der{\ln p}{\ln r}-\der{\ln\rho}{\ln r},
	\end{array}
\end{equation*}
where $g=GM_r/r^2$ is the local gravitational acceleration and $M_r=\int_0^r 4\pi\rho r^2\mathrm{d}r$. Then, after cumbersome calculations, the coupling coefficient takes the form \cite{Dziembowski1982}
\begin{widetext}
	\begin{align}
		\widetilde{\mathcal{H}}\equiv\frac{\mathcal{H}}{GM/R^3}=Z_{\alpha\beta\gamma}\int_0^1 &\left\{-\sum_k\left(A^*y_{1,k}+V_gz_k\right)\left(\varpi_{k'}\varpi_{k''}y_{1,k'}y_{1,k''}+\frac{QC_k}{c_1^2}y_{2,k'}y_{2,k''}\right)\right. \notag \\
		&+\frac{V_g}{c_1}\left[\left(V-2V_g-\der{\ln\Gamma_1}{\ln r}\right)\prod_k z_k+A_g\prod_k\left(y_{1,k}-z_k\right)\right] \notag \\
		&+\frac{A^*}{c_1}\left[\left(V_g+U-4-c_1\sum_k\varpi_k^2\right)\prod_k y_{1,k}-V_g\sum_k z_k y_{1,k'}y_{1,k''}+\sum_k y_{4,k}y_{1,k'}y_{1,k''}\right] \notag \\
		&\left.+\frac{A^*}{c_1^2}\sum_k y_{2,k}\left(GC_k y_{1,k'}y_{1,k''}+QC_{k'}y_{1,k'}z_{k''}+QC_{k''}y_{1,k''}z_{k'}\right)\right\}\rho R^5 x^4\mathrm{d} x. \label{coupling coefficient}
	\end{align}
\end{widetext}
In the expression above, the index $k$ successively takes one of the values $(\alpha,\beta,\gamma)$, whereas the indices $k'$ and $k''$ take the values that come next and after next, respectively (for example, for $k=\alpha$, $k'=\beta$ and $k''=\gamma$). The rest of the quantities are defined as
\begin{gather*}
	z_k=y_{2,k}-y_{3,k}, \\[1em]
	\varpi_k=\left\{ 
	\begin{array}{l}
		\tilde{\omega}_k \\
		-\tilde{\omega}_k 
	\end{array} 
	\right. \textrm{for} \quad
	\begin{array}{l}
		k=\alpha \\
		k=\beta,\gamma,
	\end{array} \\[1em]
	QC_k=\frac{-\Lambda_k+\Lambda_{k'}+\Lambda_{k''}}{2\varpi_{k'}\varpi_{k''}}, \\[1em]
	GC_k=\frac{\Lambda_k\varpi_k+\left(\Lambda_{k'}-\Lambda_{k''}\right)\left(\varpi_{k'}-\varpi_{k''}\right)}{2\varpi_k\varpi_{k'}\varpi_{k''}},
\end{gather*}
with $\Lambda_k=l_k\left(l_k+1\right)$. Also,
\begin{equation*}
	A_g=-\der{\ln\Gamma_1}{\ln r}-V_g\left(\parder{\ln\Gamma_1}{\ln\rho}\right)_\mu.
\end{equation*}
Finally,
\begin{equation*}
	Z_{\alpha\beta\gamma}=\iint Y_\alpha^* Y_\beta Y_\gamma\sin\theta\mathrm{d}\theta\mathrm{d}\phi,
\end{equation*}
where $Y_k\equiv Y_{l_k}^{m_k}$.

\Cref{coupling coefficient} is invariant to the transformations
\begin{equation*}
	Y_\alpha\rightleftarrows Y_\beta, \quad y_{i,\alpha}\rightleftarrows y_{i,\beta}, \quad Y_\gamma\rightarrow Y_\gamma^*, \quad \tilde{\omega}_\gamma\rightarrow -\tilde{\omega}_\gamma
\end{equation*}
and
\begin{equation*}
	Y_\alpha\rightleftarrows Y_\gamma, \quad y_{i,\alpha}\rightleftarrows y_{i,\gamma}, \quad Y_\beta\rightarrow Y_\beta^*, \quad \tilde{\omega}_\beta\rightarrow -\tilde{\omega}_\beta,
\end{equation*}
which proves that $F_{\alpha\beta\gamma}=F_{\beta\bar{\gamma}\alpha}=F_{\gamma\alpha\bar{\beta}}\equiv\mathcal{H}$.

The expression above is the zeroth-order component of the coupling coefficient, namely, all quantities are evaluated in the nonrotating limit. A more general expression could be found if we had replaced the rotationally corrected eigenfunctions in \cref{mode alpha coupling coefficient}, but this would significantly complicate the calculation.

$\mathcal{H}$ has units of energy; the normalization in \cref{coupling coefficient} is useful when all quantities in the amplitude equations of motion \eqref{equations of motion} [or \eqref{equations of motion with velocity}] are normalized accordingly. Defining a dimensionless time $\tau=t\sqrt{GM/R^3}$ and a dimensionless frequency $\tilde{\omega}=\omega/\sqrt{GM/R^3}$, the equations of motion are written
\begin{subequations}
	\label[subequations]{equations of motion normalised}
	\begin{align}
		Q'_\alpha & =\tilde{\gamma}_\alpha Q_\alpha+\frac{i\widetilde{\mathcal{H}}}{\tilde{b}_\alpha}Q_\beta Q_\gamma e^{-i\Delta\tilde{\omega}\tau}, \label{mode alpha equation of motion normalised} \\
		Q'_\beta & =\tilde{\gamma}_\beta Q_\beta+\frac{i\widetilde{\mathcal{H}}}{\tilde{b}_\beta} Q_\gamma^* Q_\alpha e^{i\Delta\tilde{\omega}\tau}, \label{mode beta equation of motion normalised} \\
		Q'_\gamma & =\tilde{\gamma}_\gamma Q_\gamma+\frac{i\widetilde{\mathcal{H}}}{\tilde{b}_\gamma} Q_\alpha Q_\beta^* e^{i\Delta\tilde{\omega}\tau}, \label{mode gamma equation of motion normalised}
	\end{align}
\end{subequations}
where $\tilde{\gamma}=\gamma/\sqrt{GM/R^3}$, $\tilde{b}=b/\sqrt{GM/R^3}$ and the prime denotes differentiation with respect to $\tau$.


\section{Study of a three-mode network with quadratic nonlinearities} \label[appsec]{sec:Study of a three-mode network with quadratic nonlinearities}


\subsection{The multiscale method} \label[appsec]{subsec:The multiscale method}

Let us assume that we have an ordinary differential equation which includes a small parameter $\epsilon$. We write the solution to this equation in the form of an asymptotic series, in the sense that
\begin{equation*}
	y(t)\rightarrow\sum_{n=0}^\infty y_n(t)\epsilon^n.
\end{equation*}
In the beginning of the evolution, when $t$ is small, low-order terms dominate the solution. However, as $t$ grows bigger, the contribution of higher-order terms cannot be neglected. These terms are usually called \emph{secular terms}, because their effects become important (compared to low-order terms) at later stages of the evolution. This behavior appears, for example, in a damped harmonic oscillator, where the zeroth-order solution is simply an undamped harmonic oscillation, with the damping effects occurring at higher orders.

The multiscale method (cf. for instance, \refcite{NayfehMook1979}) is a way to capture such higher-order effects from secular terms and make them appear in the low-order terms. As a result, the low-order approximation of the solution would be valid on secular time scales.

We define the time scales $T_n=\epsilon^n t$ and rewrite the asymptotic solution, so that
\begin{equation*}
	y(t)\rightarrow\sum_{n=0}^\infty y_n(T_0,T_1,T_2,\ldots)\epsilon^n.
\end{equation*}
In other words, we let the terms of the series depend on more than one time scale. As we will see, this allows us to ``eliminate'' secular effects from higher-order terms, thus preventing these terms from becoming significant.

We are going to use this method, in order to study \cref{equations of motion}. First, we remove the exponential time dependence by setting $C_k=Q_k\exp(i\omega_k t)$ ($k=\alpha,\beta,\gamma$) and the equations of motion are written as
\begin{subequations}
	\label[subequations]{equations of motion with no exponential time dependence}
	\begin{align}
		\dot{C}_\alpha-i\omega_\alpha C_\alpha & =\gamma_\alpha C_\alpha+\frac{i\mathcal{H}}{b_\alpha}C_\beta C_\gamma, \label{mode alpha equation of motion with no exponential time dependence} \\
		\dot{C}_\beta-i\omega_\beta C_\beta & =\gamma_\beta C_\beta+\frac{i\mathcal{H}}{b_\beta} C^*_\gamma C_\alpha, \label{mode beta equation of motion with no exponential time dependence} \\
		\dot{C}_\gamma-i\omega_\gamma C_\gamma & =\gamma_\gamma C_\gamma+\frac{i\mathcal{H}}{b_\gamma} C_\alpha C^*_\beta. \label{mode gamma equation of motion with no exponential time dependence}
	\end{align}
\end{subequations}
Now, we seek solutions of the form
\begin{equation*}
	C_k=\epsilon C_k^{(1)}(T_0,T_1)+\epsilon^2 C_k^{(2)}(T_0,T_1)+\ldots,
\end{equation*}
where $T_0=t$ and $T_1=\epsilon t$. Time derivatives then become
\begin{equation*}
	\der{}{t}=\parder{}{T_0}+\der{T_1}{T_0}\parder{}{T_1}=\parder{}{T_0}+\epsilon\parder{}{T_1}.
\end{equation*}
Replacing the solutions in \cref{equations of motion with no exponential time dependence} and distinguishing between $\mathcal{O}(\epsilon)$ and $\mathcal{O}(\epsilon^2)$ terms, we get
\begin{align*}
	\parder{C_\alpha^{(1)}}{T_0}-i\omega_\alpha C_\alpha^{(1)}=0, \\
	\parder{C_\beta^{(1)}}{T_0}-i\omega_\beta C_\beta^{(1)}=0, \\
	\parder{C_\gamma^{(1)}}{T_0}-i\omega_\gamma C_\gamma^{(1)}=0,
\end{align*}
and
\begin{align*}
	\parder{C_\alpha^{(1)}}{T_1}+\parder{C_\alpha^{(2)}}{T_0}-i\omega_\alpha C_\alpha^{(2)} & =\hat{\gamma}_\alpha C_\alpha^{(1)}+\frac{i\mathcal{H}}{b_\alpha} C_\beta^{(1)} C_\gamma^{(1)}, \\
	\parder{C_\beta^{(1)}}{T_1}+\parder{C_\beta^{(2)}}{T_0}-i\omega_\beta C_\beta^{(2)} & =\hat{\gamma}_\beta C_\beta^{(1)}+\frac{i\mathcal{H}}{b_\beta} C_\gamma^{*(1)} C_\alpha^{(1)}, \\
	\parder{C_\gamma^{(1)}}{T_1}+\parder{C_\gamma^{(2)}}{T_0}-i\omega_\gamma C_\gamma^{(2)} & =\hat{\gamma}_\gamma C_\gamma^{(1)}+\frac{i\mathcal{H}}{b_\gamma} C_\alpha^{(1)} C_\beta^{*(1)},
\end{align*}
respectively, where we also set $\gamma_k=\epsilon\hat{\gamma}_k$, so that damping and nonlinear terms appear in the same order.

The first-order equations have simple solutions of the form
\begin{equation}
	C_k^{(1)}(T_0,T_1)=A_k(T_1) e^{i\omega_k T_0}, \label{amplitude solutions first order}
\end{equation}
which we substitute to the second-order equations, to get
\begin{widetext}
	\begin{subequations}
	\label[subequations]{equations of motion second-order terms}
		\begin{align}
			\parder{C_\alpha^{(2)}}{T_0}-i\omega_\alpha C_\alpha^{(2)} & =\left(\hat{\gamma}_\alpha A_\alpha-\der{A_\alpha}{T_1}\right)e^{i\omega_\alpha T_0}+\frac{i\mathcal{H}}{b_\alpha}A_\beta A_\gamma e^{i(\omega_\beta+\omega_\gamma)T_0}, \label{mode alpha equation of motion second-order terms} \\
			\parder{C_\beta^{(2)}}{T_0}-i\omega_\beta C_\beta^{(2)} & =\left(\hat{\gamma}_\beta A_\beta-\der{A_\beta}{T_1}\right)e^{i\omega_\beta T_0}+\frac{i\mathcal{H}}{b_\beta} A_\gamma^* A_\alpha e^{i(\omega_\alpha-\omega_\gamma)T_0}, \label{mode beta equation of motion second-order terms} \\
			\parder{C_\gamma^{(2)}}{T_0}-i\omega_\gamma C_\gamma^{(2)} & =\left(\hat{\gamma}_\gamma A_\gamma-\der{A_\gamma}{T_1}\right)e^{i\omega_\gamma T_0}+\frac{i\mathcal{H}}{b_\gamma} A_\alpha A_\beta^* e^{i(\omega_\alpha-\omega_\beta)T_0}. \label{mode gamma equation of motion second-order terms}
		\end{align}
	\end{subequations}
\end{widetext}
As we mentioned earlier, the whole point of the multiscale method is to transfer long-term effects from higher-order terms to low-order terms. In this case, we want to prevent the second-order terms of the solution, $C_k^{(2)}$, from growing and becoming important. To accomplish this, we have to eliminate the so-called secular terms. In the case of \cref{equations of motion second-order terms}, terms that include the factor $\exp{(i\omega_k T_0)}$ have to vanish, because they produce secular terms, causing the solution to grow in time.


\subsubsection{The nonresonant case} \label{subsubsec:The nonresonant case}

If there is no resonance of the form $\omega_\alpha\approx\omega_\beta+\omega_\gamma$ between the modes, then the conditions for the elimination of secular terms from \cref{equations of motion second-order terms} are
\begin{equation*}
	\der{A_k}{T_1}=\hat{\gamma}_k A_k,
\end{equation*}
or
\begin{equation*}
	A_k=a_k e^{\hat{\gamma}_k T_1},
\end{equation*}
which makes the first-order solutions \eqref{amplitude solutions first order}
\begin{equation*}
	C_k=\epsilon C_k^{(1)}+\mathcal{O}(\epsilon^2)=\epsilon a_k e^{\gamma_k t} e^{i\omega_k t}+\mathcal{O}(\epsilon^2),
\end{equation*}
or, in terms of the original variables $Q_k$,
\begin{equation}
	Q_k=\epsilon a_k e^{\gamma_k t}+\mathcal{O}(\epsilon^2). \label{amplitude solutions for the non-resonant case}
\end{equation}
\Cref{amplitude solutions for the non-resonant case} shows that, if there is no resonance between the modes, their amplitudes grow or decrease with time, depending on the sign of $\gamma_k$.


\subsubsection{The resonant case} \label{subsubsec:The resonant case}

If a resonance of the form $\omega_\alpha=\omega_\beta+\omega_\gamma+\Delta\omega$ exists ($\Delta\omega$ being a small detuning), then the second terms on the right-hand sides of \cref{equations of motion second-order terms} also contribute in the production of secular terms in the solution. Then, the secular-term elimination conditions become
\begin{subequations}
	\label[subequations]{solvability conditions}
	\begin{align}
		\der{A_\alpha}{T_1} & =\hat{\gamma}_\alpha A_\alpha+\frac{i\mathcal{H}}{b_\alpha} A_\beta A_\gamma e^{-i\Delta\hat{\omega}T_1}, \label{mode alpha solvability condition} \\
		\der{A_\beta}{T_1} & =\hat{\gamma}_\beta A_\beta+\frac{i\mathcal{H}}{b_\beta} A_\gamma^* A_\alpha e^{i\Delta\hat{\omega}T_1}, \label{mode beta solvability condition} \\
		\der{A_\gamma}{T_1} & =\hat{\gamma}_\gamma A_\gamma+\frac{i\mathcal{H}}{b_\gamma} A_\alpha A_\beta^* e^{i\Delta\hat{\omega}T_1}, \label{mode gamma solvability condition}
	\end{align}
\end{subequations}
where we set $\Delta\omega=\epsilon\Delta\hat{\omega}$. From \cref{solvability conditions} we obtain our original system \eqref{equations of motion}, whose study is presented in \cref{subsec:Parametric resonance instability,subsec:Equilibrium solution,subsec:Saturation conditions}.


\subsection{Linear stability analysis} \label[appsec]{subsec:Linear stability analysis}

Having used the variable transformation \eqref{amplitude and phase variables} to the equations of motion \eqref{equations of motion}, we obtain \cref{equations of motion with amplitude and phase variables}, namely
\begin{align*}
	\dot{\varepsilon}_\alpha=\gamma_\alpha\varepsilon_\alpha+\varepsilon_\beta\varepsilon_\gamma\sin\varphi, \\
	\dot{\varepsilon}_\beta=\gamma_\beta\varepsilon_\beta-\varepsilon_\gamma\varepsilon_\alpha\sin\varphi, \\
	\dot{\varepsilon}_\gamma=\gamma_\gamma\varepsilon_\gamma-\varepsilon_\alpha\varepsilon_\beta\sin\varphi,
\end{align*}
and
\begin{equation*}
	\dot{\varphi}=\cot\varphi\left[\frac{\dot{\varepsilon}_\alpha}{\varepsilon_\alpha}+\frac{\dot{\varepsilon}_\beta}{\varepsilon_\beta}+\frac{\dot{\varepsilon}_\gamma}{\varepsilon_\gamma}-\gamma\right]+\Delta\omega,
\end{equation*}
where $\varphi=\vartheta_\alpha-\vartheta_\beta-\vartheta_\gamma+\Delta\omega t$ and $\gamma=\gamma_\alpha+\gamma_\beta+\gamma_\gamma$.

We linearize these equations by imposing small perturbations around their equilibrium solutions \eqref{amplitude and phase variables equilibria}. Denoting these perturbations by $\delta$ (not to be confused with a Eulerian perturbation), we get \cite{Dziembowski1982}
\begin{subequations}
	\label[subequations]{equations of motion with amplitude and phase variables perturbed}
	\begin{align}
		\der{}{t}\left(\frac{\delta\varepsilon_\alpha}{\varepsilon_\alpha}\right) & =-\gamma_\alpha\left(-\frac{\delta\varepsilon_\alpha}{\varepsilon_\alpha}+\frac{\delta\varepsilon_\beta}{\varepsilon_\beta}+\frac{\delta\varepsilon_\gamma}{\varepsilon_\gamma}+\kappa\delta\varphi\right), \label{mode alpha equation of motion with amplitude and phase variables perturbed} \\
		\der{}{t}\left(\frac{\delta\varepsilon_\beta}{\varepsilon_\beta}\right) & =-\gamma_\beta\left(\frac{\delta\varepsilon_\alpha}{\varepsilon_\alpha}-\frac{\delta\varepsilon_\beta}{\varepsilon_\beta}+\frac{\delta\varepsilon_\gamma}{\varepsilon_\gamma}+\kappa\delta\varphi\right), \label{mode beta equation of motion with amplitude and phase variables perturbed} \\
		\der{}{t}\left(\frac{\delta\varepsilon_\gamma}{\varepsilon_\gamma}\right) & =-\gamma_\gamma\left(\frac{\delta\varepsilon_\alpha}{\varepsilon_\alpha}+\frac{\delta\varepsilon_\beta}{\varepsilon_\beta}-\frac{\delta\varepsilon_\gamma}{\varepsilon_\gamma}+\kappa\delta\varphi\right), \label{mode gamma equation of motion with amplitude and phase variables perturbed}
	\end{align}
	and
	\begin{equation}
		\der{\delta\varphi}{t}=\kappa\sum_k\varGamma_k\frac{\delta\varepsilon_k}{\varepsilon_k}+\gamma\delta\varphi, \label{phase equation of motion perturbed}
	\end{equation}
\end{subequations}
where $\kappa=\Delta\omega/\gamma$ and $\varGamma_k=2\gamma_k-\gamma$, with the index $k$ successively taking the values $(\alpha,\beta,\gamma)$.

The matrix of the linear system \eqref{equations of motion with amplitude and phase variables perturbed} is
\begin{equation*}
	\bm{A}=\left(
	\begin{array}{rrrr}
		\gamma_\alpha          & -\gamma_\alpha        & -\gamma_\alpha         & -\kappa\gamma_\alpha \\
		-\gamma_\beta          & \gamma_\beta          & -\gamma_\beta          & -\kappa\gamma_\beta  \\
		-\gamma_\gamma         & -\gamma_\gamma        & \gamma_\gamma          & -\kappa\gamma_\gamma \\
		\kappa\varGamma_\alpha & \kappa\varGamma_\beta & \kappa\varGamma_\gamma & \gamma_{\phantom{a}}
	\end{array}
	\right),
\end{equation*}
with the help of which we can find the system's characteristic polynomial, via the relation $\left|\bm{A}-\lambda\bm{I}\right|=0$, where $\lambda$ are the eigenvalues of $\bm{A}$ and $\bm{I}$ is the identity matrix. The polynomial has the form $\lambda^4+a_1\lambda^3+a_2\lambda^2+a_3\lambda+a_4=0$, where
\begin{gather*}
		a_1=-2\gamma, \quad a_2=\gamma^2\left(1+\kappa^2\right)-4\kappa^2\sum_k\gamma_k\gamma_{k'}, \\
		a_3=4\left(1+3\kappa^2\right)\prod_k\gamma_k, \quad a_4=-4\left(1+\kappa^2\right)\gamma\prod_k\gamma_k,
\end{gather*}
with the index $k'$ taking the value that comes after $k$'s value (e.g., if $k=\alpha$, $k'=\beta$).

Now, we can use the Routh-Hurwitz stability criteria (cf. for instance, \refcite{HornJohnson1991}), in order to determine the behavior of the system. First, we construct the Routh-Hurwitz matrix, using the polynomial coefficients, which is
\begin{equation*}
	\bm{M}=\left(
	\begin{array}{cccc}
		a_1 & 1   & 0   & 0   \\
		a_3 & a_2 & a_1 & 1   \\
		0   & a_4 & a_3 & a_2 \\
		0   & 0   & 0   & a_4
	\end{array}
	\right).
\end{equation*}
Then, the stability criteria are given by
\begin{align}
	W_1 & \equiv a_1>0, \label{stability criterion 1} \\
	W_2 & \equiv \left|
		\begin{array}{cc}
			a_1 & 1 \\
			a_3 & a_2
		\end{array}
	\right|=a_1a_2-a_3>0, \label{stability criterion 2} \\
	W_3 & \equiv \left|
	\begin{array}{ccc}
		a_1 & 1   & 0   \\
		a_3 & a_2 & a_1 \\
		0   & a_4 & a_3
	\end{array}
	\right|=a_3W_2-a_1^2a_4>0, \label{stability criterion 3}
\end{align}
and
\begin{equation}
	\hspace{-3.1cm} W_4\equiv\left|\bm{M}\right|=a_4W_3>0. \label{stability criterion 4}
\end{equation}

Since $\gamma_{\beta,\gamma}<0$, it can be easily shown that the second and fourth criteria are redundant and follow from the other ones. Indeed, if $W_1>0$ then $a_4$ is also positive, which, combined with $W_3>0$, makes the fourth criterion true. Also, $W_3>0$ yields $W_2>a_1^2a_4/a_3$, but since $a_3>0$, the second criterion is also true.

So, finally, from the first and third criteria we obtain the stability conditions \eqref{saturation condition 1} and \eqref{saturation condition 2}, which are further studied in \cref{subsec:Saturation conditions}.

%
\bibliography{references.bib}
%
\end{document}